\documentclass[8.5pt]{article}
\usepackage[super,sort&compress,comma]{natbib}
\usepackage{times,mathptmx,color}
\usepackage{amssymb}

\usepackage{sectsty}
\usepackage{balance}
\usepackage{subcaption}
\usepackage{graphicx}
\usepackage{lastpage}
\usepackage[format=plain,justification=raggedright,singlelinecheck=false,font=small,labelfont=bf,labelsep=space]{caption}
\usepackage{epstopdf}
\usepackage{amsmath}

\graphicspath{{Figures/}}

\newcommand{\en}{\varepsilon^n}

\newcommand{\exw}{\varepsilon_x^w}

\definecolor{darkgreen}{RGB}{0,139,0}
\definecolor{turqoise}{RGB}{64,224,208}
\definecolor{brown}{RGB}{210,105,30}

\definecolor{b}{rgb}{0,0,1.0}
\definecolor{r}{rgb}{1,0,0}
\definecolor{g}{rgb}{0,1,0}

\begin{document}

\thispagestyle{plain}
\renewcommand{\thefootnote}{\fnsymbol{footnote}}
\renewcommand\footnoterule{\vspace*{1pt}%
\hrule width 3.4in height 0.4pt \vspace*{5pt}}
\setcounter{secnumdepth}{5}

\makeatletter
\def\subsubsection{\@startsection{subsubsection}{3}{10pt}{-1.25ex plus -1ex minus -.1ex}{0ex plus 0ex}{\normalsize\bf}}
\def\paragraph{\@startsection{paragraph}{4}{10pt}{-1.25ex plus -1ex minus -.1ex}{0ex plus 0ex}{\normalsize\textit}}
\renewcommand\@biblabel[1]{#1}
\renewcommand\@makefntext[1]%
{\noindent\makebox[0pt][r]{\@thefnmark\,}#1}
\makeatother
\renewcommand{\figurename}{\small{Fig.}~}
\sectionfont{\large}
\subsectionfont{\normalsize}

\setlength{\arrayrulewidth}{1pt}
\setlength{\columnsep}{6.5mm}
\setlength\bibsep{1pt}

\newcommand{\rev}[1]{{\color{red}#1}}
\twocolumn[
  \begin{@twocolumnfalse}
\noindent\LARGE{\textbf{Programmable Mechanical Metamaterials: the Role of Geometry} \textit{$^{\dagger}$}}
\vspace{0.6cm}

\noindent\large{\textbf{Bastiaan Florijn,\textit{$^{a,b}$} Corentin Coulais,\textit{$^{a,b}$} and
Martin van Hecke\textit{$^{a,b}$}}}\vspace{0.5cm}

\noindent\textit{\small{\textbf{Received Xth XXXXXXXXXX 20XX, Accepted Xth XXXXXXXXX 20XX\newline
First published on the web Xth XXXXXXXXXX 200X}}}

\noindent \textbf{\small{DOI: 10.1039/b000000x}}
\vspace{0.6cm}

We experimentally and numerically study the precise role of geometry for the mechanics of biholar metamaterials, quasi-2D slabs of rubber patterned by circular holes of two alternating sizes. We recently showed how the response to uniaxial compression of these metamaterials can be programmed by their lateral confinement \cite{2014FlorijnPRL}. In particular, there is a range of confining strains $\varepsilon_x$ for which the resistance to compression becomes non-trivial - non-monotonic or hysteretic - in a range of compressive strains $\varepsilon_y$. Here we show how the dimensionless geometrical parameters $t$ and $\chi$, which characterize the porosity and size ratio of the holes that pattern these metamaterials, can significantly tune these ranges over a wide range. We study the behavior for the limiting cases where $t$ and $\chi$ become large, and discuss the new physics that arises there. Away from these extreme limits, the variation of the strain ranges of interest is smooth with porosity, but the variation with size ratio evidences a cross-over at low $\chi$ from biholar to monoholar (equal sized holes) behavior, related to the elastic instabilities in purely monoholar metamaterials \cite{Bertoldi_AdvM2010}. Our study provides precise guidelines for the rational design of programmable biholar metamaterials, tailored to specific applications, and indicates that the widest range of programmability arises for moderate values of both $t$ and $\chi$.
\vspace{0.5cm}
 \end{@twocolumnfalse}
  ]

\footnotetext{\textit{$^{a}$~Huygens-Kamerling Onnes Lab, Universiteit Leiden, P.O. Box 9504, 2300 RA, Leiden, The Netherlands }}
\footnotetext{\textit{$^{b}$~FOM Institute AMOLF, Science Park 104, 1098 XG Amsterdam, The Netherlands }}
\footnotetext{\textit{$^{\dagger}$~Electronic supplementary information (ESI) available. See DOI: }}
\section{Introduction}

Mechanical metamaterials derive their unusual properties from their architecture, rather than from their composition~\cite{Wegener_reviewRPP2008}. The essentially unlimited design space of architectures thus opens up the opportunity for rational design of designer materials~\cite{EMLHecke}, functional forms of  matter with carefully crafted properties. Precise geometric design has resulted in metamaterials with negative Poisson's ratio~\cite{Lakes_science1987}, negative compressibility~\cite{Lakes_Nature2001,Nicolaou_NATMAT2012}, tunable ratio of shear to bulk modulus~\cite{Milton_JMPS1992,Kadic_APL2012,Buckmann_Natcomm2014,sidPRL} and topological nontrivial behavior~\cite{Paulose23062015,Chen09092014,Kane2014}. Going beyond linear response, a range of metamaterials have been developed which harness geometric nonlinearities and elastic instabilities to obtain novel functionalities, such as pattern switching \cite{Mullin_PRL2007,Bertoldi_JMPS2008,Bertoldi_AdvM2010,Overvelde_AdvM2012,Shim_SM2013} and sequential shape changes \cite{overveldePNAS,TangAdvMat}.\\
A currently emerging theme is the use of frustration to obtain more complex behavior, including multistability \cite{WaitukaitisPRL,LechenaultPRL,Silverberg647}. We recently showed how to leverage frustration and prestress in soft mechanical metamaterials
to obtain a (re)pro\-gram\-mable mechanical response~\cite{2014FlorijnPRL}. These metamaterials are quasi-2D slabs of rubber, patterned with a square array of circular holes of alternating sizes $D_1$ and $D_2$ (Fig.~1). In these biholar samples, one of the $90^{\circ}$ rotational symmetries, present for equal hole sizes ($D_1=D_2$), is broken, and as a consequence, the deformations patterns corresponding to purely horizontal ($x$) or vertical ($y$) compression are distinct. This sets up a competition when the material first is confined in the lateral $x$-direction, and then is uniaxially compressed in the $y$-direction with strain $\varepsilon_y$ and corresponding force $F_y$. Indeed, we found that the mechanical response $F_y(\varepsilon_y)$ can be tuned qualitatively by varying the lateral confinement $\varepsilon_x$. In particular we showed that depending on $\varepsilon_x$, the material could exhibit a non-monotonic response, where $\partial_{\varepsilon_y} F_y <0$ for a range of vertical strains, as well as a hysteretic response where $F_y(\varepsilon_y)$ becomes multi-valued~\cite{2014FlorijnPRL}.

\begin{figure}[t]
        \centering
        \begin{subfigure}[b]{0.23\textwidth}
								\caption{}
								\vspace{-0.3cm}
                \includegraphics[width=1\textwidth, trim={1cm 1cm 2cm 0cm},clip]{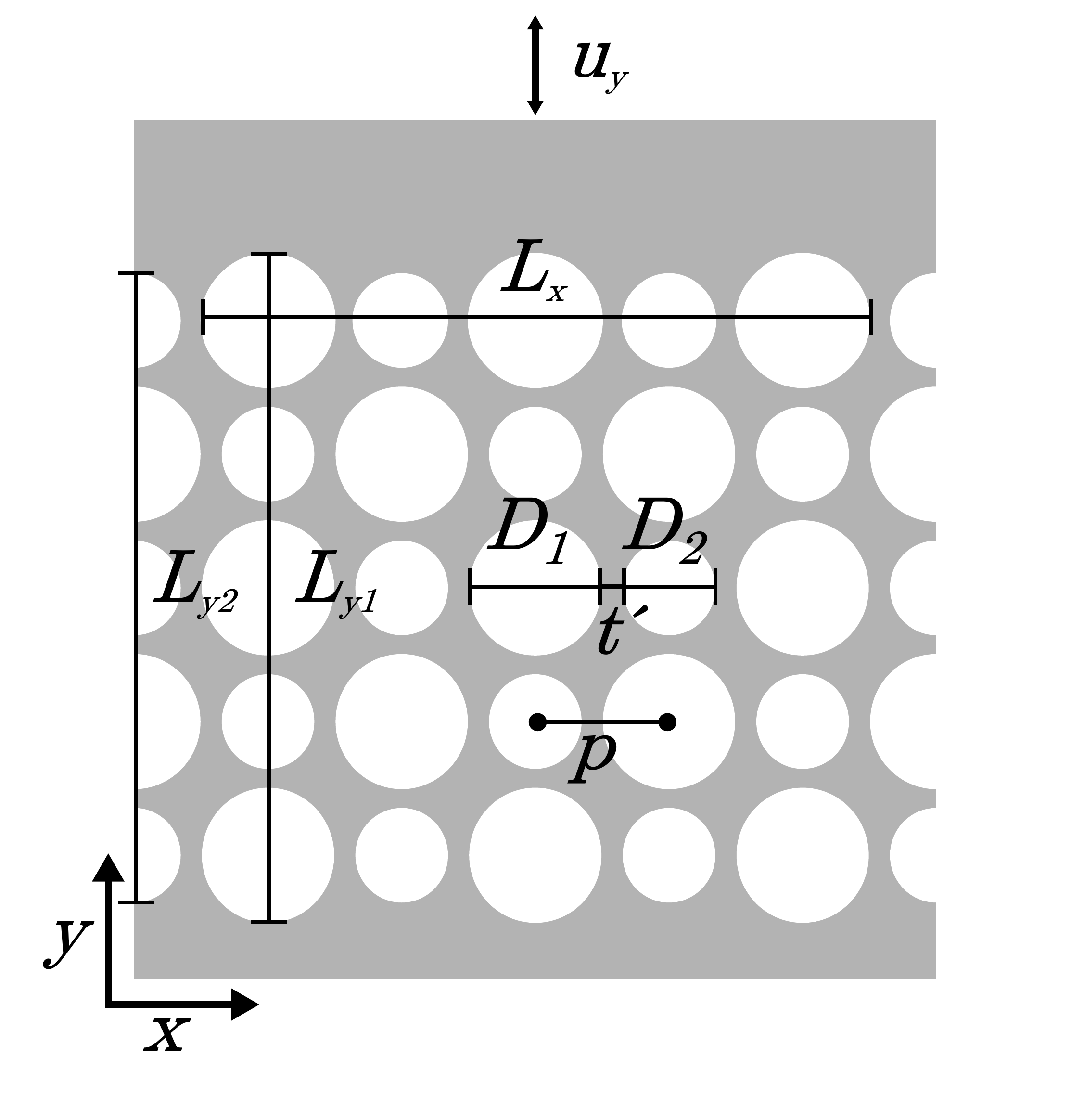}
								\label{parameters}

        \end{subfigure}
        ~
        \begin{subfigure}[b]{0.23\textwidth}
								\caption{}
								\vspace{-0.3cm}
                \includegraphics[width=1\textwidth, trim={1cm 1cm 2cm 0cm},clip]{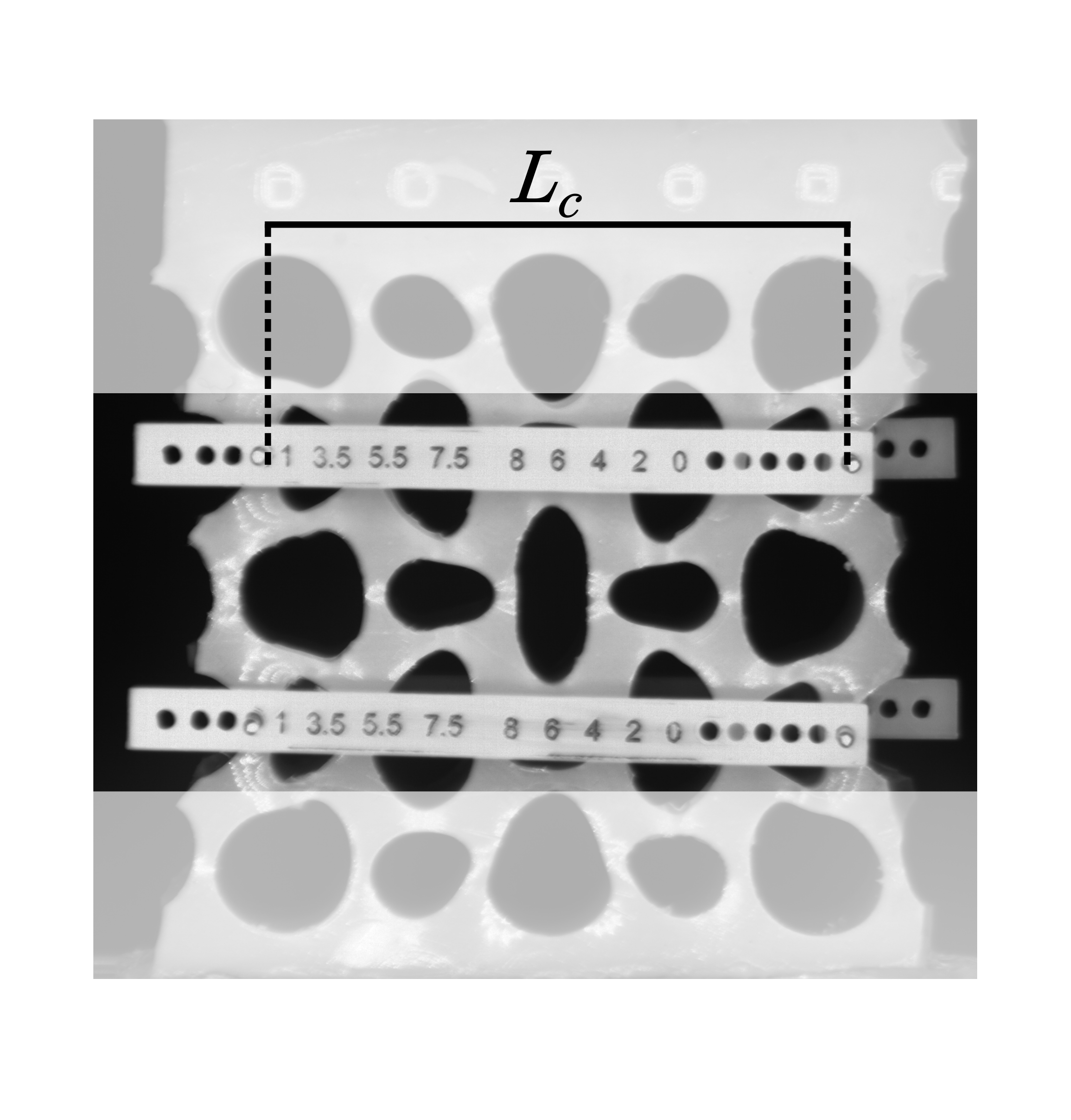}
                \label{confined}
        \end{subfigure}
				
\caption{(a) Geometry of biholar samples; $D_1$ and $D_2$ denote the hole diameters, $p$ their distance, and $t'$ the thinnest part of the filaments. The region of interest is characterized by $L_{x}$, $L_{y1}$ and $L_{y2}$.
(b) Horizontally confined sample. $L_c$ denotes the distance between the confining pins}
\label{fig:locallength}
\end{figure}

\noindent
Here we study the generality of these findings by varying the thickness of the elastic  filaments $t$ as well as the degree of biholarity $\chi$, i.e. the size difference between small and large holes. We start by showing that fully 3D numerical simulations capture the experimental findings, and allow to distinguish truly hysteretic behavior from minor visco-elastic effects inevitably present in the polymer samples.
We introduce order parameters to identify and classify the transitions between monotonic, non-monotonic and hysteretic behavior, and probe their scaling near the regime transitions. We then scan the design parameter space and show that programmable behavior persists for a wide range of the geometrical parameters $t$ and $\chi$. Moreover, we formulate design strategies to strongly tune the range of vertical strains where behavior of interest, i.e., non-monotonic or hysteretic response arises. Finally, we explore extreme limits of these design parameters, and find that most useful behavior occurs for moderate values --- the limits of large and small $t$ and $\chi$ all lead to new instabilities or singular behavior that hinder functionality. Our study thus opens a pathway to the rational, geometrical design of programmable biholar metamaterials, tailored to exhibit non-monotonic or hysteretic behavior for desired strain ranges.

\section{Samples and Experimental Methods}
To fabricate biholar metamaterials, we pour a two component silicone elastomer (Zhermack Elite Double 8, Young’s Modulus $E \simeq 220$ kPa, Poisson's ratio $\nu\simeq 0.5$) in a $120 \times 65 \times 35 $ mm mold, where cylinders of diameters $D_1 \geq D_2$ are alternately placed in a $5 \times 5$ square grid of pitch $p=10$ mm (the central cylinder has diameter $D_1$)\cite{2014FlorijnPRL}. To slow down cross linking, leaving time for the material to degas and fill every nook and cranny in the mold, we cool down these components to $-18^\circ$C. When the cross-linking process has finished (after approximately 1hr at room temperature) we remove the material from the mold and cut the lateral sides. We let the sample rest for one week, after which the elastic moduli have stopped aging. This results in samples with $5\times5$ holes, as shown in Fig.~\ref{fig:locallength}. All experiments are carried out for samples of thickness $d = 35$ mm, to avoid out of plane buckling. We characterize our samples by their biholarity $\chi:= (D_1-D_2)/p$ and dimensionless thickness $t:= 1- (D_1 + D_2)/(2p) = t'/p$.\\
\noindent
We glue the flat top and bottom parts of the material to two acrylic plates that facilitate clamping in our uniaxial compression device. Under compression, deformations are concentrated in the central part of the sample. We focus on this region of interest, and define the compressive vertical strain as:
\begin{equation}
\varepsilon_y=\frac{2u_y}{L_{y1}+L_{y2}+2t'}~,
\end{equation}
\noindent
where $(L_{y1}+L_{y2}+2t')/2$ is the effective size of the vertical region of interest and
$u_y$ the imposed deformation (Fig.~\ref{parameters}).\\
\noindent
To impose lateral confinement, we glue copper rods of diameter $1.2$~mm on the sides of our samples and use laser cut, perforated acrylic clamps fix to the distance $L_c$ between these rods (Fig.~\ref{confined}). Note that even and odd rows of our sample have different lateral boundaries, and we only clamp the 2nd and 4th row (Fig.~\ref{confined}). The global confining strain is $\varepsilon_x=1-L_c/L_{c0}$, with $L_{c0}$ the distance between the metal rods without clamps.\\
In our experiments, we measure the force $F$ as function of the compressive vertical strain $\varepsilon_y$. We define a
dimensionless effective stress as:
\begin{equation}
S := \frac{\sigma_{y}}{E} \frac{A_{\mbox{eff}}}{A} = \frac{6t'F}{dE(L_{x}+2t')^2}  ~,
\end{equation}
where $\sigma_{y} = F/A$, $A = d(L_{x}+2t')$ denotes the cross section, $L_{x}+2t'$ is the width of the region of interest,
$A_{\mbox{eff}} = 6t'd$ denotes the effective cross section, and $E$ the Young's Modulus.\\
\noindent
To characterize the spatial configuration, we fit an ellipse to the shape of the central hole, and define its polarization $\Omega$ as \cite{2014FlorijnPRL}:
\begin{equation}
\Omega=  \pm (1-p_2/p_1)\cos2\phi ,
\end{equation}
\noindent
where $p_1$ and $p_2$ are the major and minor axes of the ellipse, and $\phi$ is the angle between the major and $x$-axis. We fix the sign of $\Omega$ such that it is positive for samples that are predominantly compressed in the $y$-direction.\\
To uniaxially compress the sample while probing its response, we use an Instron 5965 uniaxial testing device. The device controls the vertical motion of a horizontal cross bar with a resolution of 4 $\mu$m. The sample is clamped  between a ground plate and this moving  bar, and we measure the compressive force $F$ with a $100$ N load cell with $5$ mN resolution. To calibrate force $F=0$ at $\varepsilon_y=0$ and at zero lateral confinement, we attach the unconfined sample to the top clamps, and then attach bottom and side clamps.\\
\noindent
For each experiment, we perform a strain sweep as follows: we first stretch the sample to $u_y=-
4$ mm, then compress to $u_y=8$ mm and finally decompress to $u_y = 0$ mm to complete the sweep. The deformation rate is fixed at $0.1$ mm per second: at this rate, visco-elastic and creep effects are minimal (Fig. S1a and b, ESI$^{\dagger}$). A high resolution camera ($2048 \times 2048$ pixels, Basler acA2040-25gm) acquires images of the deformed samples and tracks the positions and shapes of the holes with a spatial resolution of $0.03$ mm in order to determine the polarization and the confining strain $\varepsilon_x$. The image acquisition is synchronized with the data acquisition of the Instron device, running at a rate of 2Hz.

\section{Numerical Simulations}

\label{Biholarity:Simulations}

In parallel, we have performed a full parametric study of the role of $\chi$ and $t$ using 3D finite element simulations in \\
ABAQUS/STANDARD (version 6.13). We performed uniaxial compression simulations on a laterally confined sample with the same geometry, clamping and dimensions as in experiments using realistic, boundary conditions at the top and bottom of the sample. A horizontal confining strain is applied by fixing the $x$-coordinates of an arc of the boundary holes of every even row, similar to the experiments. The length of the arc is set constant at $S_c=1.1$ mm, which closely matches experimental conditions. (Note that the arc length has a minor influence on the mechanical response, but does not affect the overall phenomenology, Fig.~S2, ESI$^{\dagger}$.)\\
We model the rubber used in the experiments as an incompressible neo-Hookean continuum solid \cite{Coulais2015, Overvelde2014351}, with a strain energy density function \cite{Boyce2000, Ogden}:

\begin{equation}
W = \frac{\mu}{2}\left(\mbox{det} (\textbf{F})^{-\frac{2}{3}} \mbox{tr}(\textbf{F} \textbf{F}^{\dag} )-3\right) + \frac{K}{2} \left( \mbox{det} (\textbf{F})-1 \right) ^2,
\end{equation}

\noindent
where $\mu$ is the shear modulus, $K$ is the bulk modulus and $\textbf{F} = \partial \textbf{x} / \partial \textbf{X}$ is the deformation gradient tensor, with $\textbf{x}$ and $\textbf{X}$ the deformed and undeformed coordinates. A strictly incompressible material ($\nu =0.5$) can not be modeled with \\
ABAQUS/STANDARD, and we therefore choose $\nu =0.4990$ and $E = 220$ kPa, consistent with experiments. We use a 15-node quadratic triangular prism shape elements (ABAQUS type C3H15H). As we expect and observe only small deformations in the out-of-plane directions, we use two elements across the depth of the sample. We have performed a systematic mesh refinement study for the in-plane grid, leading to an optimal mesh size of $t'/2$.
\\
\noindent
We perform uniaxial compression tests on our confined samples. To numerically capture hysteresis, we follow two different paths for compression and decompression. The compression protocol matches the experimental protocol: First the top and bottom boundaries of the sample are fixed and the horizontal confining strain $\varepsilon_x$ is applied. Then, an increasing strain $\varepsilon_y$ is applied. The decompression protocol differs from the experimental protocol to allow the sample to reach to hysteresis related second branch. First, the sample is maximally compressed in the $y$-direction. Then, the horizontal confining strain $\varepsilon_x$ is applied. Finally, the vertical strain is lowered. These two distinct protocols allow to accurately capture the behavior on both branches in the case of hysteresis.

\section{Experimental and Numerical Results}
We perform uniaxial compression tests on $5 \times 5$ biholar samples for a range of horizontal confinements. In parallel we perform 3D realistic numerical simulations using the same geometries, clamping and boundary conditions. In the following we start by comparing experiments to simulations for a sample with $t=0.15$ and $\chi = 0.2$ and identify four qualitatively different mechanical responses, that we refer to as type ($i$)-($iv$)\cite{2014FlorijnPRL}. Next, we define order parameters that characterize these different regimes and allow us to pinpoint their transitions.

\subsection{Phenomenology}
\begin{figure}[h]
  \centering
    \includegraphics[width=.45\textwidth, trim={0cm 8cm 7.5cm 1.5cm},clip]{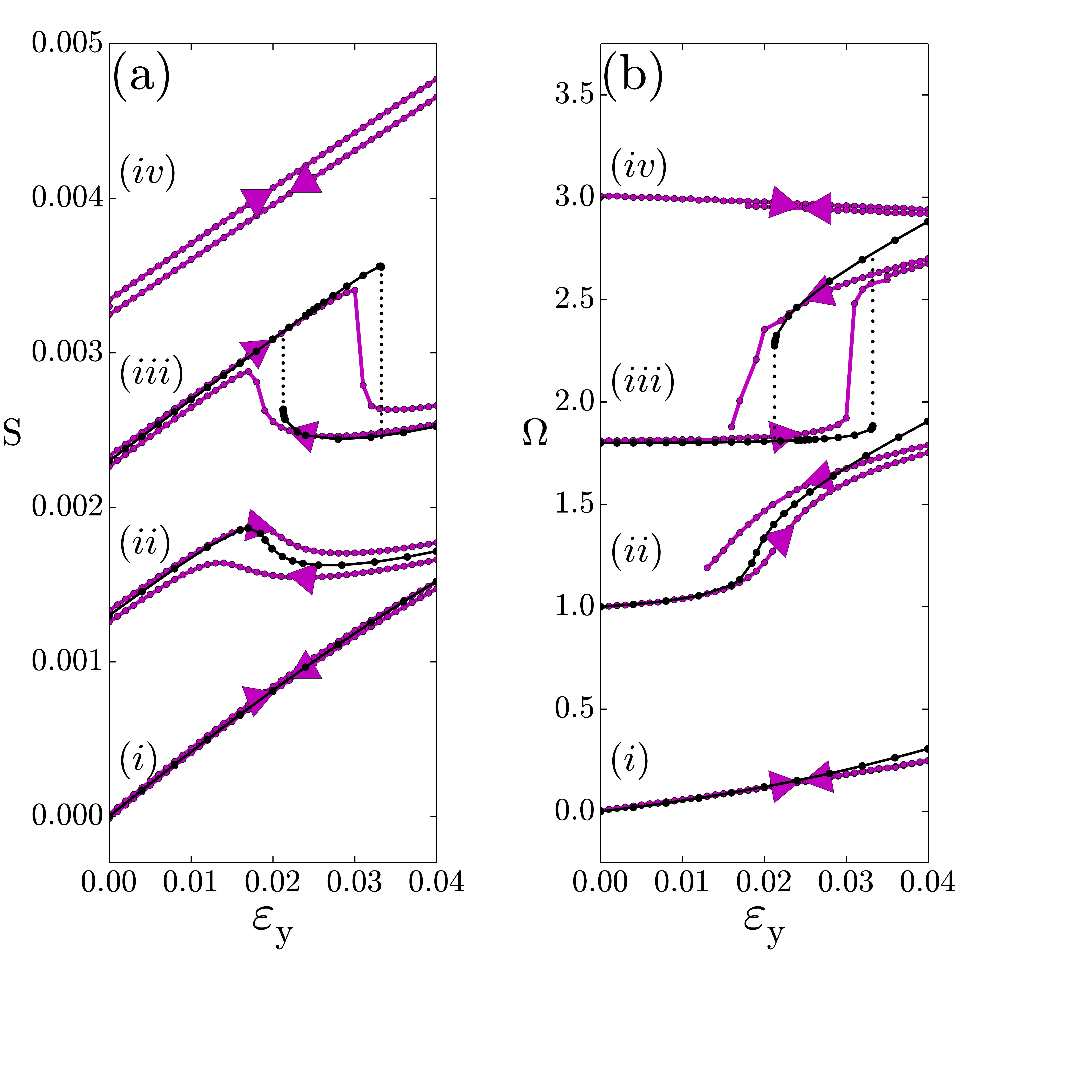}
    \caption{(a) Stress-strain curves $S(\varepsilon_y)$ for samples with $5 \times 5$ holes, $\chi = 0.2$ and $t=0.15$ (curves are offset for clarity). The horizontal confining strain $\varepsilon_x$ in curves ($i$)-($iv$) equals $\varepsilon_x = 0.000$, $0.158$, $0.178$ and $0.218$. Experimental errorbars on $\varepsilon_x$ are estimated to be $0.0025$ and are mainly caused by the manual application of the clamps. Experimental data is in magenta, and numerical data in black. (b) Corresponding plots of the polarizations $\Omega (\varepsilon_y)$ (curves are offset for clarity).}
    \label{D9p5D7p5}
\end{figure}

In Fig.~\ref{D9p5D7p5} we present the stress-strain curves, $S(\varepsilon_y)$, and pola\-ri\-zation-strain curves, $\Omega(\varepsilon_y)$, for a biholar sample with $\chi = 0.2$ and $t = 0.15$ at four different values of the horizontal confining strain. We observe a close correspondence between the numerical and experimental data, without any adjustable parameters. We distinguish four qualitatively different types of mechanical response:\\

\noindent
($i$) For small confinement, both the rescaled stress $S$ and polarization $\Omega$ increase monotonically with strain.
In experiments, both the stress and polarization exhibit a tiny amount of hysteresis. We have determined the experimental rate dependence of this hysteresis, and find that it reaches a broad minimum for the moderate rates used in the experiments, but that it increases for both very fast runs and very slow runs --- we attribute the former to viscoelastic effects,
and the latter to creep. Indeed, this residual hysteresis occurs mainly when the pattern changes rapidly, Fig.~S1c and d ESI$^{\dagger}$, and hysteresis is absent in our purely elastic numerical simulations. We conclude that non-elastic effects lead to a small hysteresis, and have adjusted our experimental rate to minimize hysteresis.\\
\noindent
 \begin{figure}[t]
        \centering
         \includegraphics[width=.45\textwidth, trim={0cm 0cm 0cm 0cm},clip]{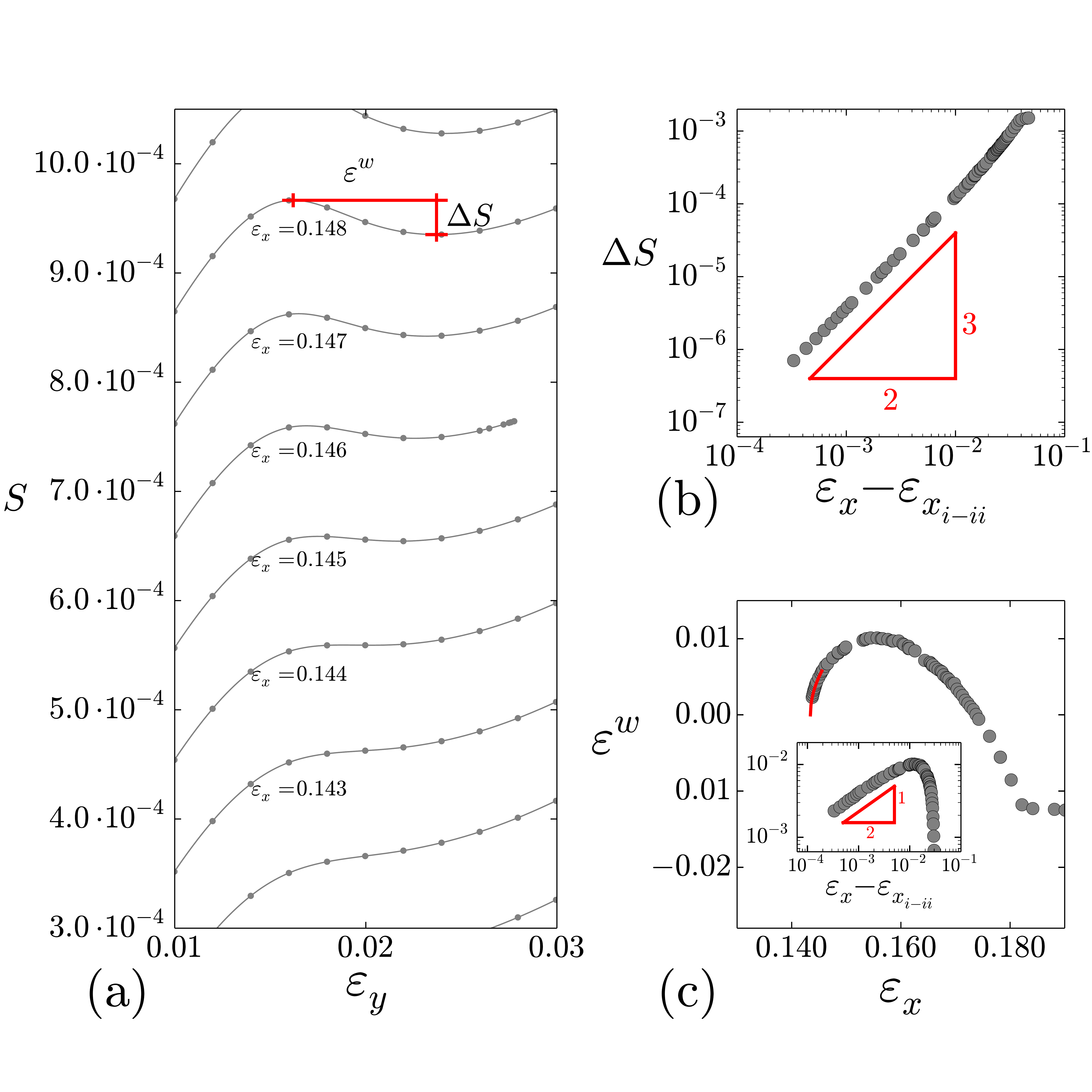}
				\caption{(a) Numerically obtained $S(\varepsilon_y)$-curves illustrating the monotonic to non-monotonic ($i$-$ii$)-transition, for a sample with $\chi = 0.2$ and $t = 0.15$ (curves offset for clarity). (b) $\Delta S$ clearly shows power law behavior, and can be fitted as $\Delta S \approx \lambda (\varepsilon_x - \varepsilon_{x_{i-ii}})^{3/2}$, where $\lambda \approx 0.117$ and $\varepsilon_{x_{i-ii}} \approx  0.143$. (c) In regime $ii$, $ \varepsilon^w $ is initially rapidly increasing and then reaches a maximum around $\varepsilon_x = 0.155$. Close to the the ($i$-$ii$)-transition, $\varepsilon^w$ shows square root behavior: $\varepsilon^w \approx \gamma (\varepsilon_x - \varepsilon_{x_{i-ii}})^{1/2}$, with $\gamma \approx 0.128$ and $\varepsilon_{x_{i-ii}} \approx 0.143$.}
         \label{i_ii_transistion}
\end{figure}
($ii$) For moderate confinement, the rescaled stress $S$ exhibits a non-monotonic increase with $\varepsilon_y$, thus featuring a range with \emph{negative incremental stiffness}. The creep-induced hysteresis in experimental data is more pronounced than in regime ($i$), but again is absent in numerical simulations (black dashed line). The polarization remains monotonic in $\varepsilon_y$, with most of its variation focused in the strain-range of negative incremental stiffness. \\
($iii$) For large confining strains, both the stress-strain curve and the polarization-strain curve exhibit a clear hysteretic transition. Away from this true hysteresis loop, the up and down sweeps are identical in simulations but differ slightly in experiments, due to the same visco-elastic effects discussed above. We note that in the numerics, the hysteretic jump between different branches is very sharp (dotted line in Fig. \ref{D9p5D7p5}), whereas in the experiments this jump is smeared out.  In the numerics, the location of the jump reproduces well, but in experiments we observe appreciable scatter between subsequent runs. We suggest that close to the jump, the system is very sensitive to imperfections, and have confirmed, by simulations, that slight geometric perturbations cause similar scatter (not shown).\\

\begin{figure}[t]
        \centering
         \includegraphics[width=.45\textwidth, trim={0cm 0cm 0cm 0cm},clip]{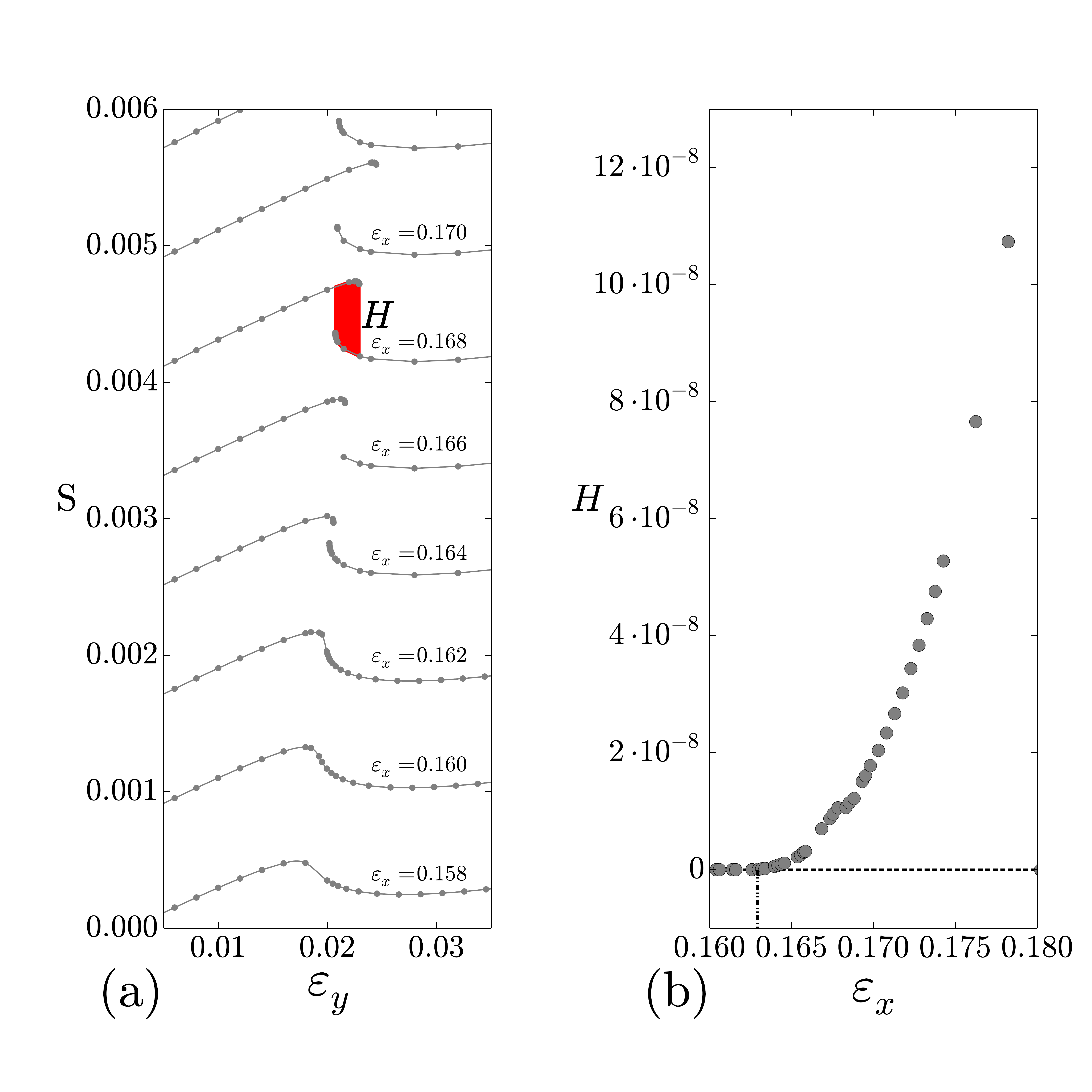}
				\caption{ (a) Numerically obtained  $S(\varepsilon_y)$-curves illustrating the non-monotonic to hysteretic $ii$-$iii$-transition, for a sample with $\chi = 0.2$ and $t = 0.15$ (curves offset for clarity). In regime ($iii$) the $S(\varepsilon_y)$-curve follow a different path for compression and decompression. The hysteresis is the area between these two paths, in the region of overlap. (b) Past the $ii$-$iii$-transition $H$ increases rapidly, with $\varepsilon_x = 0.163$ being the first nonzero value for the hysteresis, thus indicating the $ii-iii$-transition.}
         \label{ii_iii_transistion}
\end{figure}

\noindent
($iv$) For very large confinements, the stress increases monotonically with $\varepsilon_y$, similar to regime ($i$). However, the polarization is \emph{decreasing} monotonically with $\varepsilon_y$, in contrast to regime ($i$), and $\Omega$ becomes increasingly $x$-polarized under compression. Additional experiments reveal that initial compression in the $y$-direction followed by $x$-confinement brings the material to a strongly $y$-polarized state (not shown). Hence, for strong biaxial confinement there are two stable states, the order of applying $x$-confinement and $y$-compression matters, and once in the $x$-polarized state, $y$-compression is not sufficient to push the system to the $y$-polarized state.\\
We thus observe four distinct mechanical responses in a single biholar sample, depending on the amount of lateral confinement. In addition, we find very good agreement between experiments and simulations, and in the following, we focus exclusively on numerical data, as simulations do  not suffer from creep and allow for high precision and a wide range of parameters.

\begin{figure}[t]
        \centering
         \includegraphics[width=.45\textwidth, trim={0cm 0cm 0cm 0cm},clip]{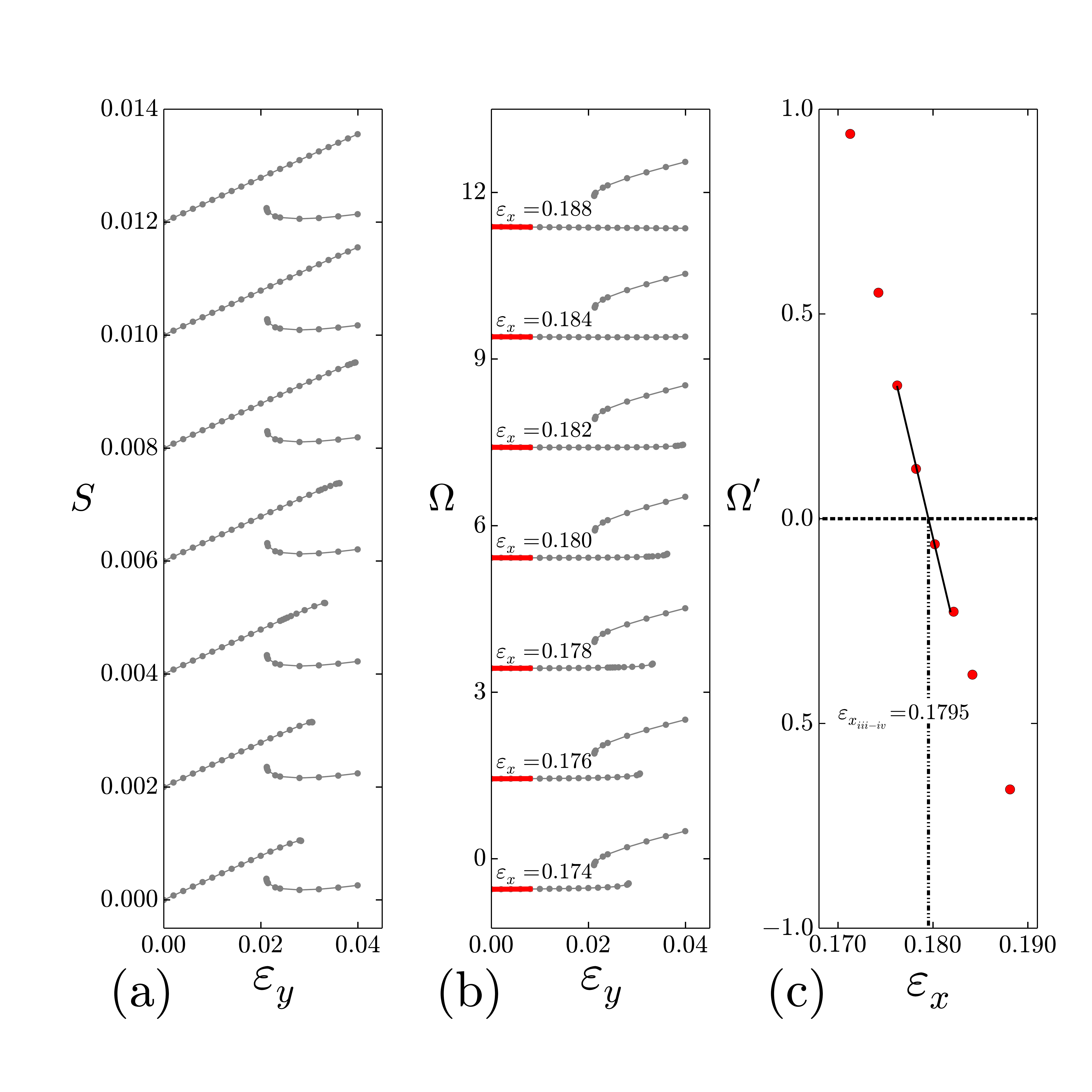}
				\caption{(a) A series of $S(\varepsilon_y)$-curves across the  hysteretic to monotonic ($iii$-$iv$)-transition, for a sample with $\chi = 0.2$ and $t = 0.15$ (curves offset for clarity). (b) The series of corresponding $\Omega(\varepsilon_y)$-curves, illustrating the $iii$-$iv$-transition. Highlighted in red the linear fit used to calculate the slope $\Omega '$. (c) Across the ($iii$-$iv$)-transition $\Omega '$ is linearly decreasing from positive values to negative values. By fitting a linear function we find, rounded off at 3 decimal digits, $\varepsilon_{x_{iii-iv}} = 0.180$.}
         \label{iii_iv_transistion}
\end{figure}

\subsection{Order Parameters}

To study whether the same scenario involving regimes ($i-iv$) is also observed for different geometries, and to investigate how the transitions between these re\-gimes vary with $t$ and $\chi$, we introduce three order parameters that allow the detection of these regimes and their transitions.

\subsubsection{($i$-$ii$)-transition}:\
Depicted in Fig.~\ref{i_ii_transistion}a is a series of $S(\varepsilon_y)$-curves illustrating the transition between monotonic and non-monotonic behavior.
In principle the sign of the incremental stiffness $\partial S / \partial \varepsilon_y$ distinguishes between these, but as the incremental stiffness is a differential quantity, a more robust measure is produced by the (existence of) local maxima and minima, which we use to determine the difference in stress, $\Delta S$, and strain, $\varepsilon^w$ (see Fig.~\ref{i_ii_transistion}a).\\
\noindent
In Fig.~\ref{i_ii_transistion}b we present $\Delta S$ as a function of the confining strain $\varepsilon_x$. Notice that $\Delta S$ rapidly increases with $\varepsilon_x$ in regime ($ii$) (and ($iii$)). The variation of $S(\varepsilon_y)$ with $\varepsilon_x$ suggest that near the transition, $S(\varepsilon_y,\varepsilon_x)$ can be expanded as: $S(\varepsilon_y) \approx \alpha (\varepsilon_x - \varepsilon_{x_{i-ii}})\varepsilon_y + \beta \varepsilon_y^3$, where $\varepsilon_{x_{i-ii}}$ is the critical horizontal strain at the ($i$-$ii$)-transition and $\alpha$ and $\beta$ are constants. We therefore expect that $\Delta S \approx (\varepsilon_x - \varepsilon_{x_{i-ii}})^{3/2}$, which is consistent with the data when we take $\varepsilon_{x_{i-ii}} = 0.143$ (Fig.~\ref{i_ii_transistion}b).
 \\
In Fig.~\ref{i_ii_transistion}c we show the strain range of negative incremental stiffness, $\varepsilon^w$, as a function of confining strain $\varepsilon_x$. Like $\Delta S$, $\varepsilon^w$ is undefined for monotonic curves, and increases rapidly with $\varepsilon_x$. As expected from our expansion of $S(\varepsilon_y)$, close to the ($i$-$ii$)-transition, we find power law scaling: $\varepsilon^w \approx (\varepsilon_x - \varepsilon_{x_{i-ii}})^{1/2}$, with the same estimate for $\varepsilon_{x_{i-ii}}$ as before, see Fig.~\ref{i_ii_transistion}c. For larger $\varepsilon_x$, $\varepsilon^w$ is decreasing and eventually becomes negative, which signals the approach to the hysteretic regime.  \\

\begin{figure}[t]
        \centering
         \includegraphics[width=.45\textwidth, trim={3cm 4cm 1cm 6cm},clip]{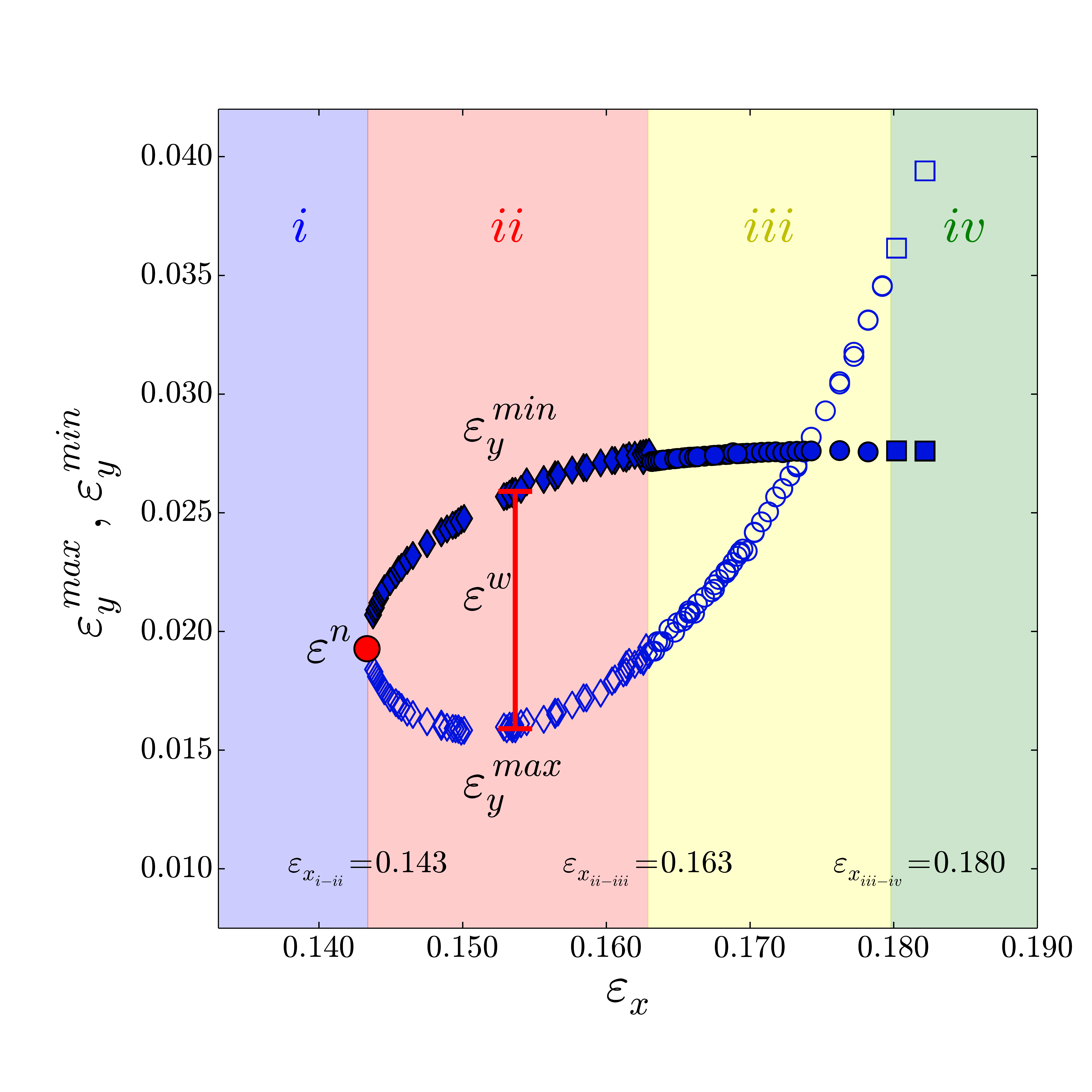}
				\caption{Representation of the characteristic strains for a sample with $\chi = 0.2$ and $t=0.15$. The red circle indicates the 'nose', ($\varepsilon^n_x,\varepsilon^n_y$), which signals the onset of regime ($ii$). Non monotonic behavior in regime ($ii$) occurs for strains between $\varepsilon_y^{min}$ (closed diamonds) and $\varepsilon_y^{max}$ (open diamonds). We extend these minimum and maximum into regime ($iii$) (circles) and regime ($iv$) (squares). The width between the two branches $\varepsilon_y^{max}$ and $\varepsilon_y^{min}$ determines the order parameter $\varepsilon^w$. The transitions between ($ii$)- ($iii$) and ($iii$) - ($iv$) cannot be detected from $\varepsilon_y^{min}$ and $\varepsilon_y^{max}$ alone and we use $H$ to detect the onset of regime ($iii$) and $\Omega$ to detect the onset of regime ($iv$).}
         \label{Fishch0p2t0p15}
\end{figure}

\begin{figure*}[t]
        \centering
         \includegraphics[width=.95\textwidth, trim={0cm 0cm 0cm 0cm},clip]{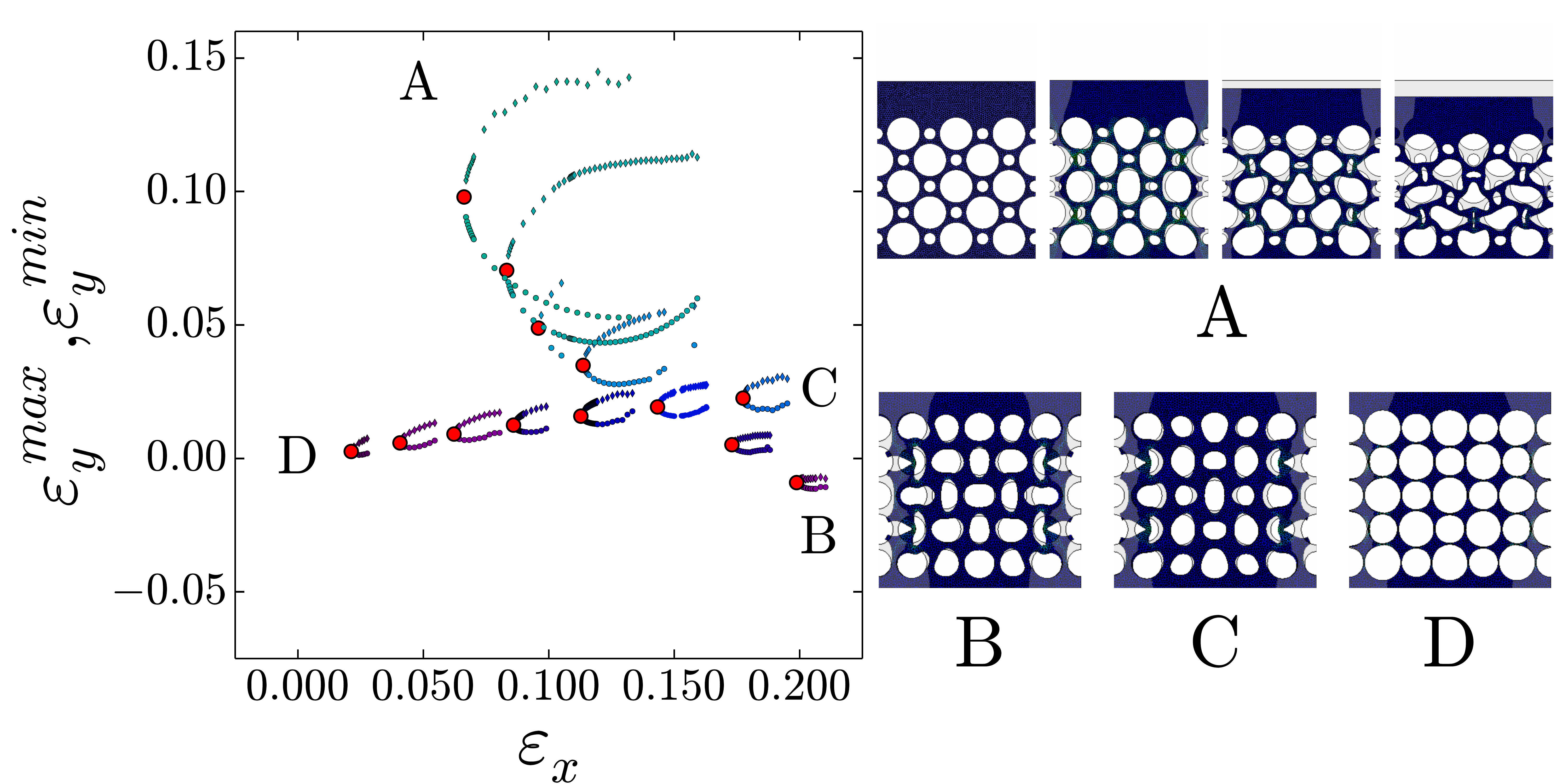}
				\caption{Strain at the local maximum $\varepsilon_y^{max}$ (circles) and local minimum $\varepsilon_y^{min}$ (diamonds) for data obtained in regime ($ii$) as a function of horizontal confinement $\varepsilon_x$ for a samples with different geometries. The red dot indicates the 'nose' of the curves. The nearly horizontal red dots correspond to $\chi=0.2$ and (from left to right) $t =0.025$, $0.050$, $0.075$, $0.100$, $0.125$, $0.150$, $0.175$, whereas the diagonally order range of red dots correspond to $t=0.15$ and (top to bottom) $\chi =0.6, 0.5, 0.4, 0.3, 0.2, 0.15$ and $0.125$. The labels $A-D$ indicate to large or small $t$ or $\chi$ limits where new behavior sets in as shown to the right. For large $\chi$ (A, $t=0.15$, $\chi=.8$), the deformation patterns become irregular; shown here are the outcome of simulations for $\varepsilon_x =0$ and $\varepsilon_y= 0$, and $\varepsilon_x =0.126$ and $\varepsilon_y= 0$, $0.062$ and $0.126$. For small $\chi$ (B, $t=0.15$, $\chi=0.1$, $\varepsilon_x=0.216$, $\varepsilon_y=0 $), and for large $t$ (C, $t=0.2$, $\chi=0.2$, $\varepsilon_x=0.206$, $\varepsilon_y=0$),
the confining strains required to obtain non-monotonic behavior become so large, that deformations become localized near the boundary and {\em sulcii} develop. Finally, for small $t$ (D, $t=0.025$, $\chi=0.2$, $\varepsilon_x=0.020$, $\varepsilon_y=0 $), the characteristic strains and strain ranges become vanishingly small.
  }
         \label{Fish}
\end{figure*}

\subsubsection{($ii$-$iii$)-transition}: \ We present in Fig.~\ref{ii_iii_transistion}a a number of $S(\epsilon_y)$-curves to illustrate the transition from nonmonotonic to hysteretic behavior. As discussed above, to numerically capture the hysteresis, we use two distinct protocols for compression and decompression. We quantify the amount of hysteresis by $H$, the area of the hysteresis loop. As shown in Fig.~\ref{ii_iii_transistion}b,  $H$ increases rapidly with the confining strain, which allows us to accurately determine the onset of hysteresis, the first non zero value for $H$, as $\varepsilon_{x_{ii-iii}} \approx 0.163$.

\subsubsection{($iii$-$iv$)-transition} \ As shown in Fig. \ref{iii_iv_transistion}a, we are unable to observe the $iii$-$iv$-transition from the $S(\varepsilon_y)$-curves. Therefore, we focus on the polarization $\Omega$ of the central hole of the sample, see Fig. \ref{iii_iv_transistion}b. We define the transition between regime $(iii)$ and $(iv)$ to occur when the polarization for small strain $\varepsilon_y$ has a negative slope ($\Omega ' < 0$), see Fig. \ref{iii_iv_transistion}c. Using a linear fit we find $\varepsilon_{x_{iii-iv}} \approx 0.180$. As the ($iii$-$iv$)-transition is not associated with any significant change in $S(\varepsilon_y)$, in the remainder we focus on the transitions to nonmonotic and hysteretic behavior.\\
\noindent
Using the order parameters $\Delta S$, $\varepsilon_w$, $H$ and $\Omega '$, we are now in a position to identify the nature of the mechanical response; monotonic ($i$), non-monotonic ($ii$), hysteretic ($iii$) or monotonic with decreasing polarization ($iv$).

\section{Parametric Study}
In the following we study how the vertical and horizontal strains where nonmonotonic and hysteretic behavior occurs vary with the geometrical design parameters $\chi$ and $t$. For each value of these parameters, we can in principle obtain $S(\varepsilon_y ,\varepsilon_x)$ and $\Omega(\varepsilon_y ,\varepsilon_x)$, from which we then can determine the strain-ranges corresponding to regime $(i-iv)$ using the order parameters defined above. We study this parameter space systematically using a large number of simulations. To do so, we have explored $7$ values of $\chi$ and $6$ values of $t$. For each set of these parameters, we have determined the relevant range of strains,  and performed simulations for typically 50 values of both  $\varepsilon_x$ and $\varepsilon_y$, leading to a total 
number of $10^5$ simulations.
Moreover, for the most interesting regimes $(ii-iii)$ we can calculate the range of vertical strains$^{\ddagger}$ $\varepsilon_y$  where the non-monotonic respectively hysteretic behavior takes place. However, the resulting deluge of data is difficult to visualize or interpret. In Fig.~6 we show a simple representation which captures the main features of the strain ranges of regime $(ii-iii)$, here for fixed $\chi$ and $t$. From $S(\varepsilon_y ,\varepsilon_x)$, we determine $\varepsilon_y^{max}$, $\varepsilon_y^{min}$, and $H$ as a function of $\varepsilon_x$, and plot
$\varepsilon_y^{max}$ (open symbols) and $\varepsilon_y^{min}$ (closed symbols) as a function of $\varepsilon_x$ and we use $H$ to distinguish data points in regime ($ii$) and ($iii$), and the polarization $\Omega$ to detect regime ($iv$).
In regime ($i$), $\varepsilon_y^{max}$ and $\varepsilon_y^{min}$ are not defined. The transition to regime ($ii$) corresponds to the 'nose' (red dot) of these curves (Fig.~\ref{Fishch0p2t0p15}). The representation
in Fig.~\ref{Fishch0p2t0p15} clearly shows the increase of the non-monotonic range as $\varepsilon_x$ is increased deeper into regime $(ii)$. Note that $\varepsilon_y^{max}$ and $\varepsilon_y^{min}$ cross eventually somewhere in regime ($iii$), see also Fig.~\ref{i_ii_transistion}.
As we will show, the overall trends in $\varepsilon_y^{min}$ and $\varepsilon_y^{max}$ as function of $\varepsilon_x$ are robust, with $\xi$ and $t$ setting the ``size'' and ``location'' of these fish-shaped curves.

In the remainder of this paper, we focus on regime $(ii)$, and in particular on the onset of the non-monotonic behavior as well as the maximum of $\varepsilon^{w}$. Note that all of this information can conveniently be related to the data shown in
Fig.~\ref{Fishch0p2t0p15}  --- the onset of non-monotonic behavior corresponds to the ``the nose of the fish'' at ($\varepsilon^n_x = \varepsilon_{x_{i-ii}},\varepsilon^n_y$), whereas the maximum non-monotonic range is given by $\varepsilon^{wm}$, ``the belly of the fish'', at $\varepsilon^{wm}_x$.

\begin{figure}[t!]
        \centering
         \includegraphics[width=.45\textwidth, trim={0cm 2cm 2cm 2cm},clip]{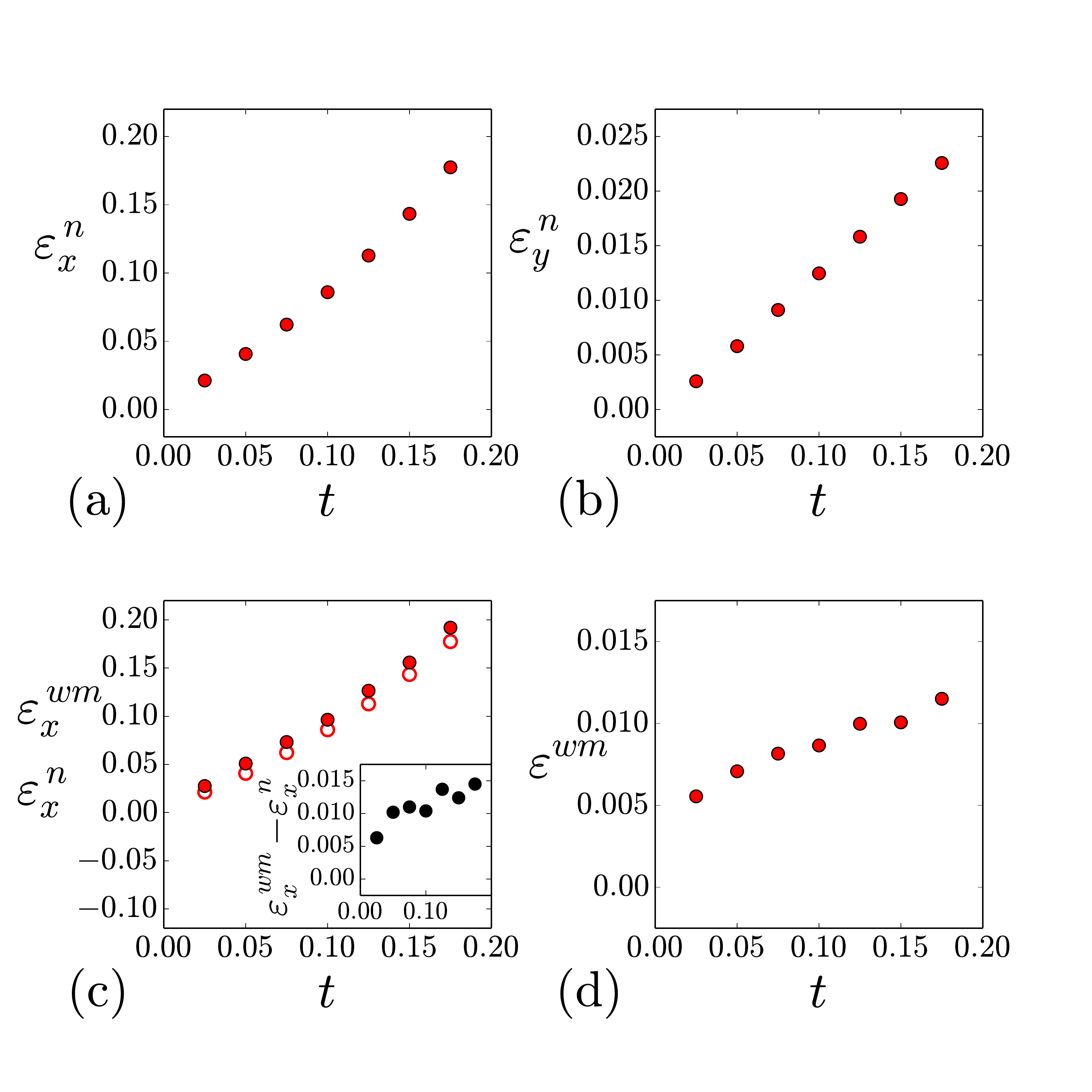}
				\caption{For fixed $\chi=0.2$, we show the variation with $\chi$ of
(a)-(b) the location  $\varepsilon^n_x$ and $\varepsilon^n_y$ of the nose which signals the transition to regime (ii), and (c-d) the $x$-location and value of the maximum difference between $\varepsilon_y^{max}$ and $\varepsilon_y^{min}$ which indicates the non-monotonic range. Black datapoints are theoretical results calculated from a biholar mechanism with $\chi = 0.2$.  }
         \label{t_trend}
\end{figure}

\begin{figure}[t!]
        \centering
         \includegraphics[width=.45\textwidth, trim={0cm 2cm 2cm 2cm},clip]{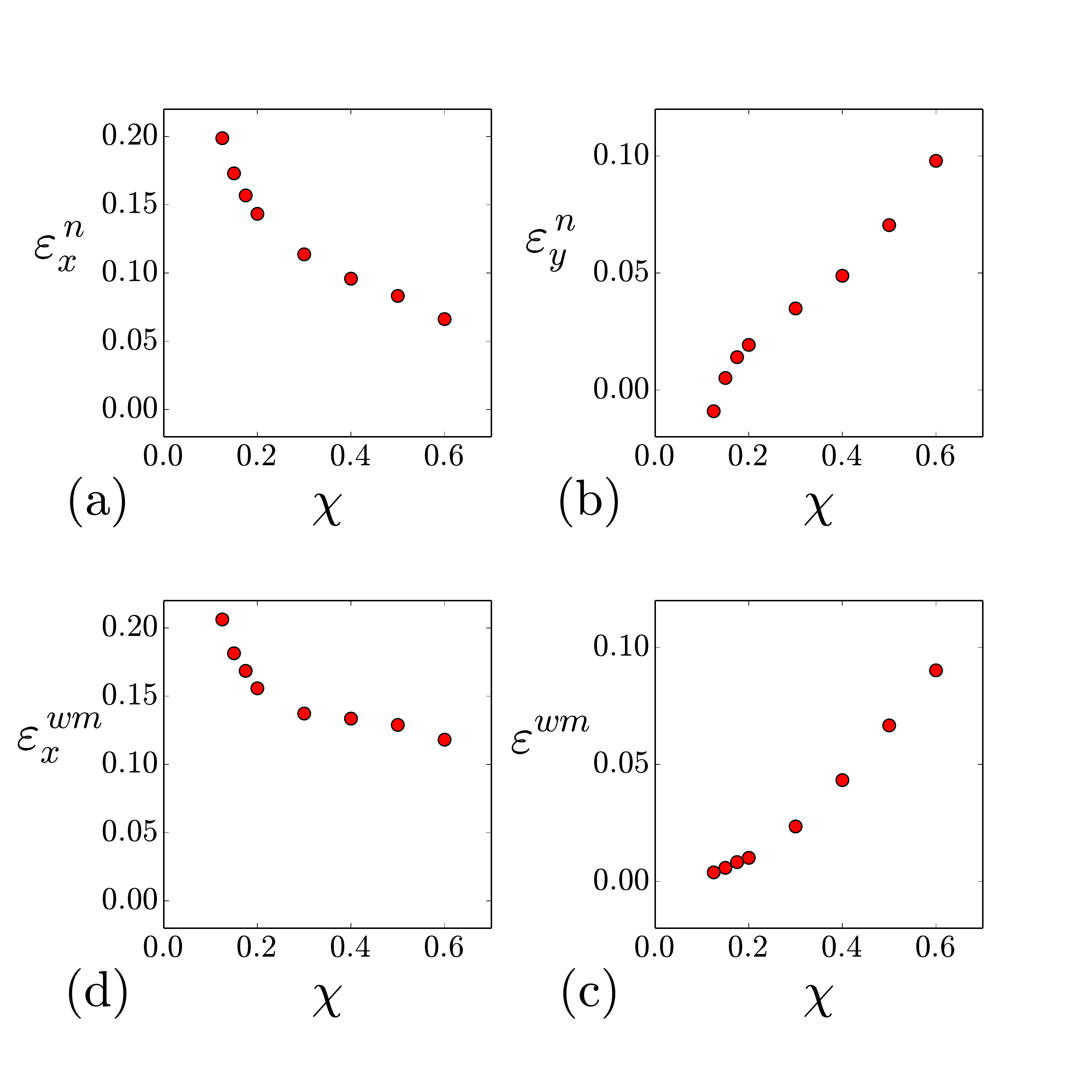}
				\caption{For fixed $t=0.15$, we show the variation with $\chi$ of
(a)-(b) the location  $\varepsilon^n_x$ and $\varepsilon^n_y$ of the nose which signals the transition to regime (ii), and (c-d) the $x$-location and value of the maximum difference between $\varepsilon_y^{max}$ and $\varepsilon_y^{min}$ which indicates the non-monotonic range. }
         \label{chi_trend}
\end{figure}

\subsection{Variation of strain ranges with geometric parameters}
\noindent
We have determined $\varepsilon_y^{max}$ and $\varepsilon_y^{min}$ for fixed $\chi=0.2$ and a range of thicknesses $t$, as well as for fixed $t=0.15$ and a range of biholarities $\chi$, as shown in Fig.~\ref{Fish}. In both cases, we can discern clear trends, as well as interesting limiting cases for large and small $t$ or $\chi$ - see Fig.~7\\
\noindent
As we vary the thickness, we observe that $\varepsilon^n_x$ and $\varepsilon^n_y$ 
smoothly decrease towards zero, whereas $\varepsilon^{wm}$ stays finite. Hence, the characteristic strains vary with $t$, but the size of the strain intervals where non-monotonic behavior occurs remains finite for small $t$.
These trends are illustrated in Fig.~\ref{t_trend}, where we show the variation of $\varepsilon^n_x$, $\varepsilon^n_y$,  $\varepsilon^{wm}$ and $\varepsilon^{wm}_x$ with $t$. In good approximation, $\varepsilon^n_x$ and $\varepsilon^n_y$ vanish linearly with $t$. As shown in Fig.~\ref{t_trend}c, even though $\varepsilon^{wm}_x$ also varies strongly with $t$, it appears to reach a finite limit for $t\rightarrow 0$, as further illustrated in the inset which shows how $\varepsilon^{wm}_x -\varepsilon^n_x$ reaches a finite value at $t=0$. Consistent with this, 
 $\varepsilon^{wm}$ approaches a finite value for $t\rightarrow 0$. 

The variation with biholarity is more significant and less simple. First, we observe that for increasing biholarity, both the
vertical and horizontal strain ranges increase significantly. Second, their typical values have opposite trends; whereas
$\varepsilon_y^{min}$ and $\varepsilon_y^{max}$ strongly increase, $\en$ and $\exw$ decrease. Hence, tuning the biholarity can be used to favor  non-monotonic behavior for small $\varepsilon_x$ or for small $\varepsilon_y$ --- including at negative vertical stresses for small values of $\chi$. Third, the range of the non-monotonic regime increases strongly with $\chi$. These trends are illustrated in Fig.~\ref{chi_trend}, where we show the variation of $\varepsilon^n_x$, $\varepsilon^n_y$,  $\varepsilon^{wm}$ and $\varepsilon^{wm}_x$ with $\chi$. This data strongly suggests that there are two distinct regimes, with a smooth crossover around $\chi \approx 0.15$. We speculate that the value of this crossover is related to $t$.
Moreover, we suggest that in the small $\chi$ regime, the materials mechanics crosses over to that of a monoholar system \cite{Mullin_PRL2007,Bertoldi_JMPS2008,Bertoldi_AdvM2010,Overvelde_AdvM2012,Shim_SM2013}, where $\varepsilon_x$ and $\varepsilon_y$ no longer are in competition and the materials behavior is difficult to program, consistent with a very small non-monotonic strain range.\\

\noindent
We can now also identify four limiting cases.
For large $\chi$, (case A in Fig.~7) we note that the small holes appear to become irrelevant, so that we approach  a monoholar system rotated by 45$^{\circ}$. In this limit, where vertical strains are large, {\em sulcii}\cite{SulcusMaha,SulcusMaha2} as well as localization bands appear \cite{Gibson}.
In the limit of vanishing $\chi$ (case B) the material approaches a monoholar material \cite{Mullin_PRL2007,Bertoldi_AdvM2010,Bertoldi_JMPS2008}, and our data suggests that these are difficult to program, with matching small non-monotonic behavior --- consistent with the absence of the broken 90$^{\circ}$ symmetry that underlies the programmability of biholar systems \cite{2014FlorijnPRL}. For small but finite $\chi$, the horizontal strains again become very large and similar as for large $t$, {\em sulcii} develop.
For large $t$ (case C), new behavior must occur --- at some point the filaments become so wide that global buckling  of the material occurs before any appreciable changes in the local pattern\cite{Overvelde2014351}. What we observe is that for large $t$ the strains needed to reach non-monotonic behavior become so large, that some of the filaments develop {\em sulcii}, so that strain localization starts to dominate the behavior --- for our systems and $\chi=0.2$, this occurs for $t>0.175$. This limits
the usefulness of large $t$ systems \cite{Bertoldi_AdvM2010}. Finally, 
in the limit of vanishing $t$ (case D), the mechanics of our system are expected to be close to the simple mechanism introduced in \cite{2014FlorijnPRL}, our numerical simulations closely match those of calculations in this model
\cite{NitinPrep}. However, here both the typical strains and strain ranges corresponding to nontrivial behavior vanish.
Hence, none of these limits are particularly useful from a practical or programmability point of view.

\section{Conclusion}

In this paper we have presented a systematic overview of the role of the geometrical design  of biholar metamaterials for obtaining reprogrammable mechanics. First, we have showed that the four qualitatively different mechanical responses $(i-iv)$ are a robust feature, and happen for a wide range of values of the design parameters $\chi$ and $t$. Second, we have identified four distinct asymptotic cases, where additional instabilities arise. Hence, programmability is optimal for moderate values of $t$ and $\chi$. Our study opens a pathway to the rational, geometrical design of programmable biholar metamaterials, tailored to exhibit non-monotonic or hysteretic behavior for desired strain ranges. Open questions for future work are to extend this frustration based strategy for the programmability of other mechanical parameters (e.g., Poissons function) \cite{Barcode} and functionalities such as tuneable damping, to smaller length scales, and to three dimensions \cite{Damping}.

\section{Acknowledgments}
We acknowledge technical assistance of Jeroen Mesman. BF, CC and MvH acknowledge funding from the Netherlands Organization for Scientific Research through a VICI grant, NWO-680-47-609.

\providecommand*{\mcitethebibliography}{\thebibliography}
\csname @ifundefined\endcsname{endmcitethebibliography}
{\let\endmcitethebibliography\endthebibliography}{}


\begin{mcitethebibliography}{33}
\providecommand*{\natexlab}[1]{#1}
\providecommand*{\mciteSetBstSublistMode}[1]{}
\providecommand*{\mciteSetBstMaxWidthForm}[2]{}
\providecommand*{\mciteBstWouldAddEndPuncttrue}
  {\def\EndOfBibitem{\unskip.}}
\providecommand*{\mciteBstWouldAddEndPunctfalse}
  {\let\EndOfBibitem\relax}
\providecommand*{\mciteSetBstMidEndSepPunct}[3]{}
\providecommand*{\mciteSetBstSublistLabelBeginEnd}[3]{}
\providecommand*{\EndOfBibitem}{}
\mciteSetBstSublistMode{f}
\mciteSetBstMaxWidthForm{subitem}
{(\emph{\alph{mcitesubitemcount}})}
\mciteSetBstSublistLabelBeginEnd{\mcitemaxwidthsubitemform\space}
{\relax}{\relax}

\bibitem[Florijn \emph{et~al.}(2014)Florijn, Coulais, and van
  Hecke]{2014FlorijnPRL}
B.~Florijn, C.~Coulais and M.~van Hecke, \emph{Physical review letters}, 2014,
  \textbf{113}, 175503\relax
\mciteBstWouldAddEndPuncttrue
\mciteSetBstMidEndSepPunct{\mcitedefaultmidpunct}
{\mcitedefaultendpunct}{\mcitedefaultseppunct}\relax
\EndOfBibitem
\bibitem[Bertoldi \emph{et~al.}(2010)Bertoldi, Reis, Willshaw, and
  Mullin]{Bertoldi_AdvM2010}
K.~Bertoldi, P.~M. Reis, S.~Willshaw and T.~Mullin, \emph{Advanced Materials},
  2010, \textbf{22}, 361--366\relax
\mciteBstWouldAddEndPuncttrue
\mciteSetBstMidEndSepPunct{\mcitedefaultmidpunct}
{\mcitedefaultendpunct}{\mcitedefaultseppunct}\relax
\EndOfBibitem
\bibitem[Kadic \emph{et~al.}(2013)Kadic, B{\"u}ckmann, Schittny, and
  Wegener]{Wegener_reviewRPP2008}
M.~Kadic, T.~B{\"u}ckmann, R.~Schittny and M.~Wegener, \emph{Rep. Prog. Phys},
  2013, \textbf{76}, 126501\relax
\mciteBstWouldAddEndPuncttrue
\mciteSetBstMidEndSepPunct{\mcitedefaultmidpunct}
{\mcitedefaultendpunct}{\mcitedefaultseppunct}\relax
\EndOfBibitem
\bibitem[Reis \emph{et~al.}(2015)Reis, Jaeger, and van Hecke]{EMLHecke}
P.~M. Reis, H.~M. Jaeger and M.~van Hecke, \emph{Extreme Mechanics Letters},
  2015, \textbf{5}, 25 -- 29\relax
\mciteBstWouldAddEndPuncttrue
\mciteSetBstMidEndSepPunct{\mcitedefaultmidpunct}
{\mcitedefaultendpunct}{\mcitedefaultseppunct}\relax
\EndOfBibitem
\bibitem[Lakes(1987)]{Lakes_science1987}
R.~Lakes, \emph{Science}, 1987, \textbf{235}, 1038--1040\relax
\mciteBstWouldAddEndPuncttrue
\mciteSetBstMidEndSepPunct{\mcitedefaultmidpunct}
{\mcitedefaultendpunct}{\mcitedefaultseppunct}\relax
\EndOfBibitem
\bibitem[Lakes \emph{et~al.}(2001)Lakes, Lee, Bersie, and
  Wang]{Lakes_Nature2001}
R.~Lakes, T.~Lee, A.~Bersie and Y.~Wang, \emph{Nature}, 2001, \textbf{410},
  565--567\relax
\mciteBstWouldAddEndPuncttrue
\mciteSetBstMidEndSepPunct{\mcitedefaultmidpunct}
{\mcitedefaultendpunct}{\mcitedefaultseppunct}\relax
\EndOfBibitem
\bibitem[Nicolaou and Motter(2012)]{Nicolaou_NATMAT2012}
Z.~G. Nicolaou and A.~E. Motter, \emph{Nature materials}, 2012, \textbf{11},
  608--613\relax
\mciteBstWouldAddEndPuncttrue
\mciteSetBstMidEndSepPunct{\mcitedefaultmidpunct}
{\mcitedefaultendpunct}{\mcitedefaultseppunct}\relax
\EndOfBibitem
\bibitem[Milton(1992)]{Milton_JMPS1992}
G.~W. Milton, \emph{Journal of the Mechanics and Physics of Solids}, 1992,
  \textbf{40}, 1105--1137\relax
\mciteBstWouldAddEndPuncttrue
\mciteSetBstMidEndSepPunct{\mcitedefaultmidpunct}
{\mcitedefaultendpunct}{\mcitedefaultseppunct}\relax
\EndOfBibitem
\bibitem[Kadic \emph{et~al.}(2012)Kadic, B{\"u}ckmann, Stenger, Thiel, and
  Wegener]{Kadic_APL2012}
M.~Kadic, T.~B{\"u}ckmann, N.~Stenger, M.~Thiel and M.~Wegener, \emph{Applied
  Physics Letters}, 2012, \textbf{100}, 191901\relax
\mciteBstWouldAddEndPuncttrue
\mciteSetBstMidEndSepPunct{\mcitedefaultmidpunct}
{\mcitedefaultendpunct}{\mcitedefaultseppunct}\relax
\EndOfBibitem
\bibitem[B{\"u}ckmann \emph{et~al.}(2014)B{\"u}ckmann, Thiel, Kadic, Schittny,
  and Wegener]{Buckmann_Natcomm2014}
T.~B{\"u}ckmann, M.~Thiel, M.~Kadic, R.~Schittny and M.~Wegener, \emph{Nature
  communications}, 2014, \textbf{5}, --\relax
\mciteBstWouldAddEndPuncttrue
\mciteSetBstMidEndSepPunct{\mcitedefaultmidpunct}
{\mcitedefaultendpunct}{\mcitedefaultseppunct}\relax
\EndOfBibitem
\bibitem[Goodrich \emph{et~al.}(2015)Goodrich, Liu, and Nagel]{sidPRL}
C.~P. Goodrich, A.~J. Liu and S.~R. Nagel, \emph{Phys. Rev. Lett.}, 2015,
  \textbf{114}, 225501\relax
\mciteBstWouldAddEndPuncttrue
\mciteSetBstMidEndSepPunct{\mcitedefaultmidpunct}
{\mcitedefaultendpunct}{\mcitedefaultseppunct}\relax
\EndOfBibitem
\bibitem[Paulose \emph{et~al.}(2015)Paulose, Meeussen, and
  Vitelli]{Paulose23062015}
J.~Paulose, A.~S. Meeussen and V.~Vitelli, \emph{Proceedings of the National
  Academy of Sciences}, 2015, \textbf{112}, 7639--7644\relax
\mciteBstWouldAddEndPuncttrue
\mciteSetBstMidEndSepPunct{\mcitedefaultmidpunct}
{\mcitedefaultendpunct}{\mcitedefaultseppunct}\relax
\EndOfBibitem
\bibitem[Chen \emph{et~al.}(2014)Chen, Upadhyaya, and Vitelli]{Chen09092014}
B.~G.-g. Chen, N.~Upadhyaya and V.~Vitelli, \emph{Proceedings of the National
  Academy of Sciences}, 2014, \textbf{111}, 13004--13009\relax
\mciteBstWouldAddEndPuncttrue
\mciteSetBstMidEndSepPunct{\mcitedefaultmidpunct}
{\mcitedefaultendpunct}{\mcitedefaultseppunct}\relax
\EndOfBibitem
\bibitem[Kane and Lubensky(2014)]{Kane2014}
C.~L. Kane and T.~C. Lubensky, \emph{Nat Phys}, 2014, \textbf{10}, 39--45\relax
\mciteBstWouldAddEndPuncttrue
\mciteSetBstMidEndSepPunct{\mcitedefaultmidpunct}
{\mcitedefaultendpunct}{\mcitedefaultseppunct}\relax
\EndOfBibitem
\bibitem[Mullin \emph{et~al.}(2007)Mullin, Deschanel, Bertoldi, and
  Boyce]{Mullin_PRL2007}
T.~Mullin, S.~Deschanel, K.~Bertoldi and M.~Boyce, \emph{Physical review
  letters}, 2007, \textbf{99}, 084301\relax
\mciteBstWouldAddEndPuncttrue
\mciteSetBstMidEndSepPunct{\mcitedefaultmidpunct}
{\mcitedefaultendpunct}{\mcitedefaultseppunct}\relax
\EndOfBibitem
\bibitem[Bertoldi \emph{et~al.}(2008)Bertoldi, Boyce, Deschanel, Prange, and
  Mullin]{Bertoldi_JMPS2008}
K.~Bertoldi, M.~Boyce, S.~Deschanel, S.~Prange and T.~Mullin, \emph{Journal of
  the Mechanics and Physics of Solids}, 2008, \textbf{56}, 2642--2668\relax
\mciteBstWouldAddEndPuncttrue
\mciteSetBstMidEndSepPunct{\mcitedefaultmidpunct}
{\mcitedefaultendpunct}{\mcitedefaultseppunct}\relax
\EndOfBibitem
\bibitem[Overvelde \emph{et~al.}(2012)Overvelde, Shan, and
  Bertoldi]{Overvelde_AdvM2012}
J.~T.~B. Overvelde, S.~Shan and K.~Bertoldi, \emph{Advanced Materials}, 2012,
  \textbf{24}, 2337--2342\relax
\mciteBstWouldAddEndPuncttrue
\mciteSetBstMidEndSepPunct{\mcitedefaultmidpunct}
{\mcitedefaultendpunct}{\mcitedefaultseppunct}\relax
\EndOfBibitem
\bibitem[Shim \emph{et~al.}(2013)Shim, Shan, Ko{\v{s}}mrlj, Kang, Chen, Weaver,
  and Bertoldi]{Shim_SM2013}
J.~Shim, S.~Shan, A.~Ko{\v{s}}mrlj, S.~H. Kang, E.~R. Chen, J.~C. Weaver and
  K.~Bertoldi, \emph{Soft Matter}, 2013, \textbf{9}, 8198--8202\relax
\mciteBstWouldAddEndPuncttrue
\mciteSetBstMidEndSepPunct{\mcitedefaultmidpunct}
{\mcitedefaultendpunct}{\mcitedefaultseppunct}\relax
\EndOfBibitem
\bibitem[Overvelde \emph{et~al.}(2015)Overvelde, Kloek, D{\textquoteright}haen,
  and Bertoldi]{overveldePNAS}
J.~Overvelde, T.~Kloek, J.~J.~A. D{\textquoteright}haen and K.~Bertoldi,
  \emph{The Proceedings of the National Academy of Sciences of the United
  States of America}, 2015, \textbf{112}, 10863--10868\relax
\mciteBstWouldAddEndPuncttrue
\mciteSetBstMidEndSepPunct{\mcitedefaultmidpunct}
{\mcitedefaultendpunct}{\mcitedefaultseppunct}\relax
\EndOfBibitem
\bibitem[Tang \emph{et~al.}(2015)Tang, Lin, Han, Qiu, Yang, and
  Yin]{TangAdvMat}
Y.~Tang, G.~Lin, L.~Han, S.~Qiu, S.~Yang and J.~Yin, \emph{Advanced Materials},
  2015, \textbf{27}, 7181--7190\relax
\mciteBstWouldAddEndPuncttrue
\mciteSetBstMidEndSepPunct{\mcitedefaultmidpunct}
{\mcitedefaultendpunct}{\mcitedefaultseppunct}\relax
\EndOfBibitem
\bibitem[Waitukaitis \emph{et~al.}(2015)Waitukaitis, Menaut, Chen, and van
  Hecke]{WaitukaitisPRL}
S.~Waitukaitis, R.~Menaut, B.~G.-g. Chen and M.~van Hecke, \emph{Phys. Rev.
  Lett.}, 2015, \textbf{114}, 055503\relax
\mciteBstWouldAddEndPuncttrue
\mciteSetBstMidEndSepPunct{\mcitedefaultmidpunct}
{\mcitedefaultendpunct}{\mcitedefaultseppunct}\relax
\EndOfBibitem
\bibitem[Lechenault and Adda-Bedia(2015)]{LechenaultPRL}
F.~Lechenault and M.~Adda-Bedia, \emph{Phys. Rev. Lett.}, 2015, \textbf{115},
  235501\relax
\mciteBstWouldAddEndPuncttrue
\mciteSetBstMidEndSepPunct{\mcitedefaultmidpunct}
{\mcitedefaultendpunct}{\mcitedefaultseppunct}\relax
\EndOfBibitem
\bibitem[Silverberg \emph{et~al.}(2014)Silverberg, Evans, McLeod, Hayward,
  Hull, Santangelo, and Cohen]{Silverberg647}
J.~L. Silverberg, A.~A. Evans, L.~McLeod, R.~C. Hayward, T.~Hull, C.~D.
  Santangelo and I.~Cohen, \emph{Science}, 2014, \textbf{345}, 647--650\relax
\mciteBstWouldAddEndPuncttrue
\mciteSetBstMidEndSepPunct{\mcitedefaultmidpunct}
{\mcitedefaultendpunct}{\mcitedefaultseppunct}\relax
\EndOfBibitem
\bibitem[Coulais \emph{et~al.}(2015)Coulais, Overvelde, Lubbers, Bertoldi, and
  van Hecke]{Coulais2015}
C.~Coulais, J.~T.~B. Overvelde, L.~A. Lubbers, K.~Bertoldi and M.~van Hecke,
  \emph{Phys. Rev. Lett.}, 2015, \textbf{115}, 044301\relax
\mciteBstWouldAddEndPuncttrue
\mciteSetBstMidEndSepPunct{\mcitedefaultmidpunct}
{\mcitedefaultendpunct}{\mcitedefaultseppunct}\relax
\EndOfBibitem
\bibitem[Overvelde and Bertoldi(2014)]{Overvelde2014351}
J.~T. Overvelde and K.~Bertoldi, \emph{Journal of the Mechanics and Physics of
  Solids}, 2014, \textbf{64}, 351 -- 366\relax
\mciteBstWouldAddEndPuncttrue
\mciteSetBstMidEndSepPunct{\mcitedefaultmidpunct}
{\mcitedefaultendpunct}{\mcitedefaultseppunct}\relax
\EndOfBibitem
\bibitem[Boyce and Arruda(2000)]{Boyce2000}
M.~C. Boyce and E.~M. Arruda, \emph{Rubber chemistry and technology}, 2000,
  \textbf{73}, 504--523\relax
\mciteBstWouldAddEndPuncttrue
\mciteSetBstMidEndSepPunct{\mcitedefaultmidpunct}
{\mcitedefaultendpunct}{\mcitedefaultseppunct}\relax
\EndOfBibitem
\bibitem[Ogden(1997)]{Ogden}
R.~W. Ogden, \emph{Nonlinear Elastic Deformations}, Dover, 1997\relax
\mciteBstWouldAddEndPuncttrue
\mciteSetBstMidEndSepPunct{\mcitedefaultmidpunct}
{\mcitedefaultendpunct}{\mcitedefaultseppunct}\relax
\EndOfBibitem
\bibitem[Hohlfeld and Mahadevan(2012)]{SulcusMaha}
E.~Hohlfeld and L.~Mahadevan, \emph{Phys. Rev. Lett.}, 2012, \textbf{109},
  025701\relax
\mciteBstWouldAddEndPuncttrue
\mciteSetBstMidEndSepPunct{\mcitedefaultmidpunct}
{\mcitedefaultendpunct}{\mcitedefaultseppunct}\relax
\EndOfBibitem
\bibitem[Hohlfeld and Mahadevan(2011)]{SulcusMaha2}
E.~Hohlfeld and L.~Mahadevan, \emph{Phys. Rev. Lett.}, 2011, \textbf{106},
  105702\relax
\mciteBstWouldAddEndPuncttrue
\mciteSetBstMidEndSepPunct{\mcitedefaultmidpunct}
{\mcitedefaultendpunct}{\mcitedefaultseppunct}\relax
\EndOfBibitem
\bibitem[Gibson and Ashby(1999)]{Gibson}
L.~J. Gibson and M.~F. Ashby, \emph{Cellular Solids: Structure and Properties
  (Cambridge Solid State Science Series)}, Cambridge University Press,
  1999\relax
\mciteBstWouldAddEndPuncttrue
\mciteSetBstMidEndSepPunct{\mcitedefaultmidpunct}
{\mcitedefaultendpunct}{\mcitedefaultseppunct}\relax
\EndOfBibitem
\bibitem[Singh \emph{et~al.}()Singh, Florijn, Coulais, and van
  Hecke]{NitinPrep}
N.~Singh, B.~Florijn, C.~Coulais and M.~van Hecke, \emph{In Preparation}\relax
\mciteBstWouldAddEndPuncttrue
\mciteSetBstMidEndSepPunct{\mcitedefaultmidpunct}
{\mcitedefaultendpunct}{\mcitedefaultseppunct}\relax
\EndOfBibitem
\bibitem[Coulais \emph{et~al.}()Coulais, Florijn, and van Hecke]{Barcode}
C.~Coulais, B.~Florijn and M.~van Hecke, \emph{In Preparation}\relax
\mciteBstWouldAddEndPuncttrue
\mciteSetBstMidEndSepPunct{\mcitedefaultmidpunct}
{\mcitedefaultendpunct}{\mcitedefaultseppunct}\relax
\EndOfBibitem
\bibitem[Coulais \emph{et~al.}()Coulais, Teomy, de~Reus, Shokef, and van
  Hecke]{Damping}
C.~Coulais, E.~Teomy, K.~de~Reus, Y.~Shokef and M.~van Hecke,
  \emph{Submitted}\relax
\mciteBstWouldAddEndPuncttrue
\mciteSetBstMidEndSepPunct{\mcitedefaultmidpunct}
{\mcitedefaultendpunct}{\mcitedefaultseppunct}\relax
\EndOfBibitem
\end{mcitethebibliography}

\footnotetext{\textit{$^{\ddagger}$~For each geometry, the resolution in $\varepsilon_y$ is set as follows. We first identify the minimal range of $\varepsilon_y$ required to observe the extrema of the  $S(\varepsilon_y ,\varepsilon_x)$-curves. This range is then divided into at least of 20 incremental static steps with additional refinements near the extrema. We then use qubic spline interpolation on each $S(\varepsilon_y ,\varepsilon_x)$-curve to measure the location of the maximum and minimum with a resolution better than $2 \cdot 10^{-4}$ of the selected strain range.}}

\clearpage

\section{Supplemental Material}
\renewcommand\thefigure{S\arabic{figure}} 
\setcounter{figure}{0}  
In the following document we provide details accompanying the paper {\em Programmable Mechanical Metamaterials: the Role of Geometry}.\\

\begin{figure}[h]
  \centering
    \includegraphics[width=.45\textwidth]{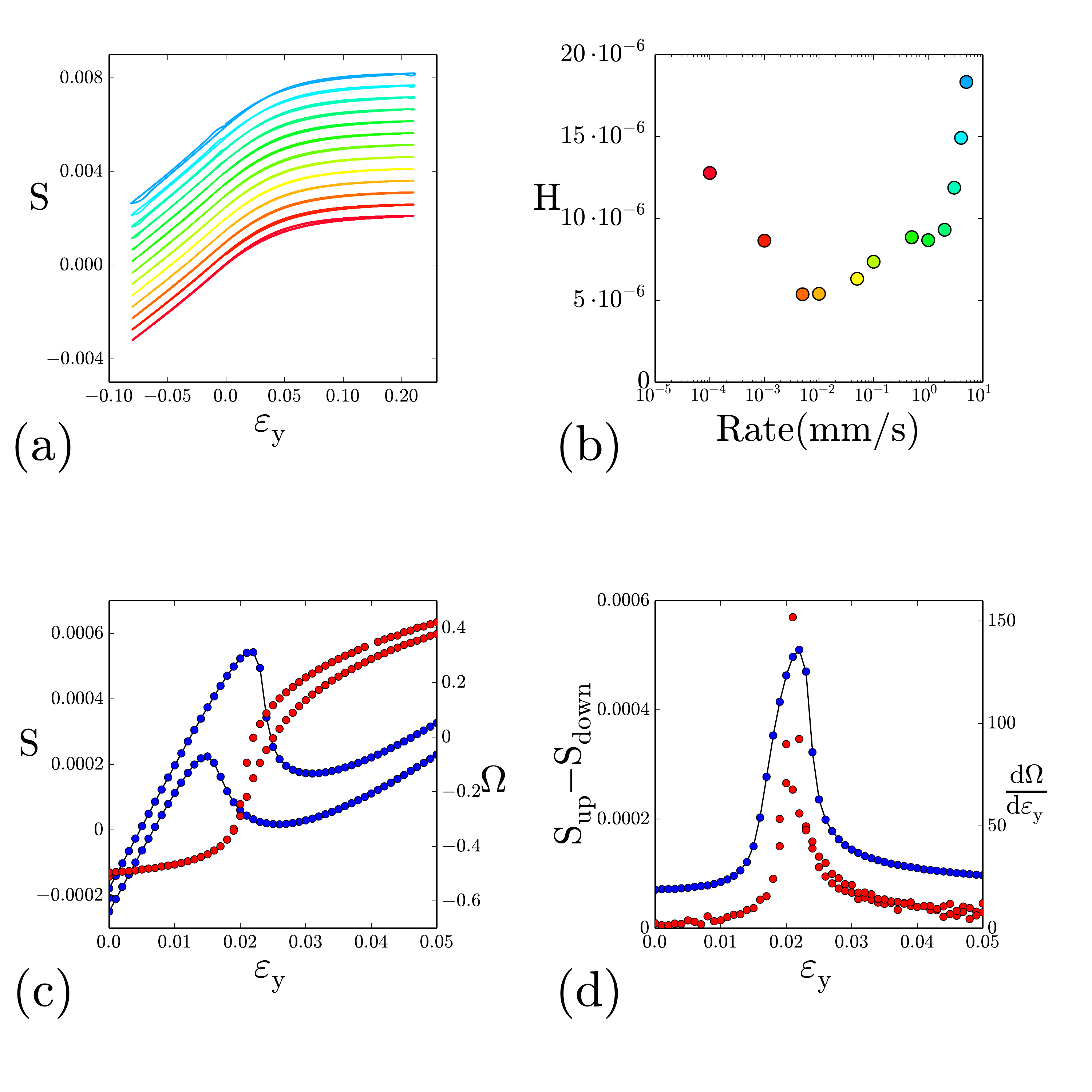}
    \caption{(a) Effective stress-strain curves for samples with $5 \times 5$ holes, dimensionless thickness $\tilde{t} = 0.15$ and biholarity $\chi =0.2$, for deformation rates varying between $10^{-4}$- $5$ mm/s (bottom to top). Curves are shown with a vertical offset for clarity. (b) Calculated hysteresis (area of the loop) as a function of deformation rate. (c) In blue the effective stress-strain curve for a samples with $5 \times 5$ holes, dimensionless thickness $\tilde{t} = 0.15$, biholarity $\chi =0.2$ and $\varepsilon_x = 0.15$ (regime $ii$), measured at a strain rate of $10^{-1}$ mm/s. In red the polarization $\Omega$ of the central hole as a function of $\varepsilon_y$.
    (d) In blue the absolute difference in $S$ during
 compression and decompression as function of $\varepsilon_y$, for the blue curve in (c). In red the derivative of the polarization $\frac{d \Omega}{ d\varepsilon_y}$, for the red curve in (c), as a function of $\varepsilon_y$. The good correspondence of the peaks in both datasets strongly suggests that hysteresis is maninly due to weak viscous effects, which are most prominent when the sample quickly changes its configuration. }
\label{HYST}
\end{figure}

\noindent
To understand the effect of the rate of deformations, we have performed a range of experiments with strain rates varying from $10^{-4}$ to $5$ mm/s (Fig.~\ref{HYST}). The results show that in a broad range of strain rates around $10^{-2}$ mm/s, spurious hysteresis is minimal. In experiments, a strain rate of $10^{-1}$ mm/s is chosen to minimize both the spurious hysteresis as well as the time it takes to perform a single experimental run. Moreover, the difference between the up and down sweep peaks when the samples quickly change their configuration, which suggest that viscous effects are responsible for this spurious hysteresis.

\begin{figure}[h]
        \centering
         \includegraphics[width=.45\textwidth, trim={0cm 0cm 0cm 0cm},clip]{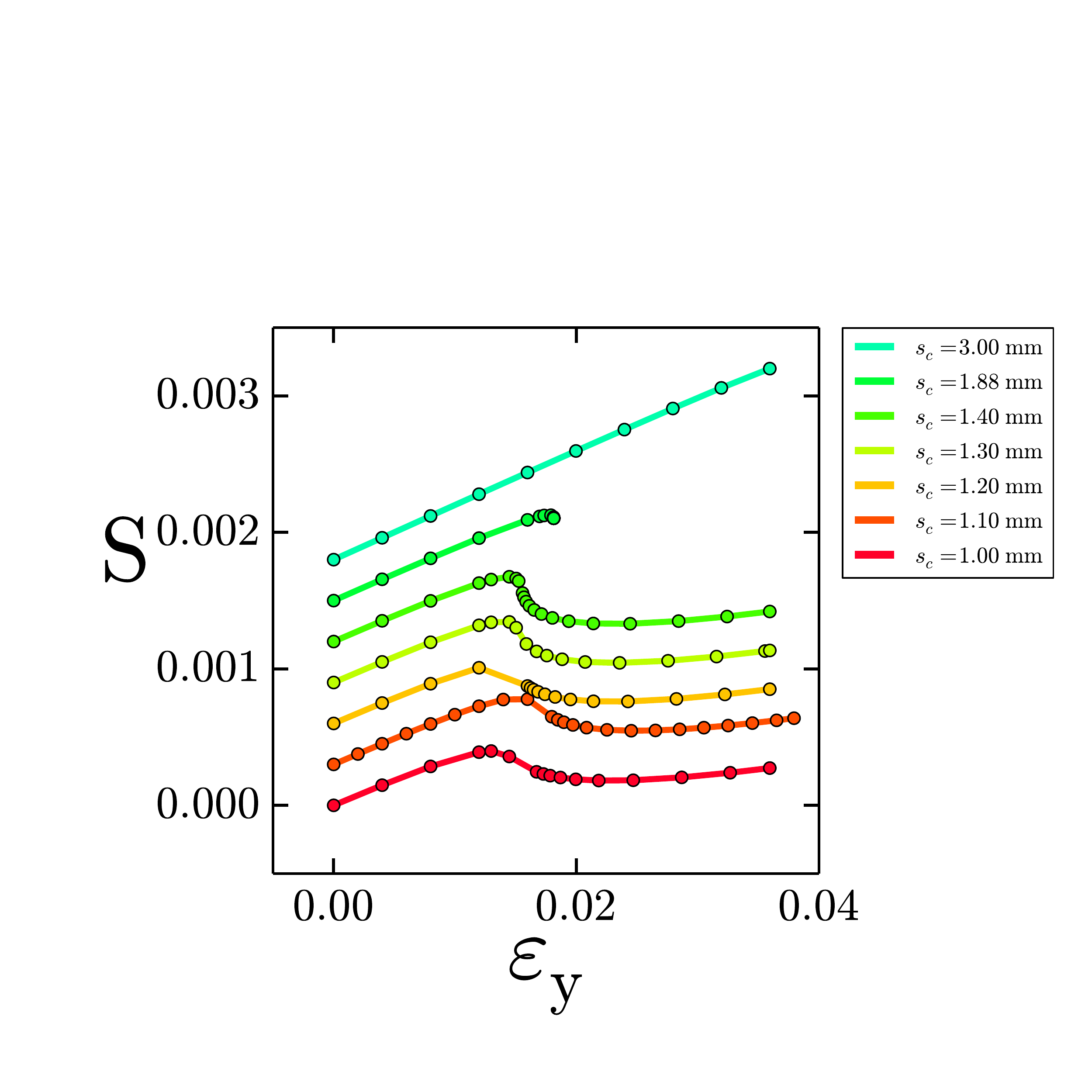}
				\caption{Numerically simulated S($\varepsilon_y$)-curves for a biholar sample with $\chi = 0.2$ and $t=0.15$ with fixed $\varepsilon_{x}= 0.1584 $. The arc length $S_c$ of the segment of the boundary holes used to confine the sample is varied; $Sc=1.10$, $1.20$, $1.30$, $1.40$, $1.88$, $3.00$.  }
         \label{clampsizes}
\end{figure}

\newpage
\noindent
In the numerical simulations we apply horizontal confining strains to our samples by fixing the $x$-coordinates of a segment of the boundary holes of every even row. In Fig.~\ref{clampsizes}, we show the effect of the arc length,$S_c$, of this segment. Increasing the arc length of the segment shifts the various regime transitions to lower values of $\varepsilon_x$.
By a comparison to our experimental data, we find that an arc length of $1.1$ mm (as used subsequently) gives the best fit, close to the actual dimension of the clamping rods used in the experiments, which have a diameter of $1.2$ mm.

\end{document}